\documentclass[a4paper,fleqn,usenatbib,useAMS]{mnras}

\usepackage{graphicx}  % Including figure files
\usepackage{amsmath}  % Advanced maths commands
\usepackage{amssymb}  % Extra maths symbols
\usepackage{multicol}        % Multi-column entries in tables
\usepackage{bm}    % Bold maths symbols, including upright Greek
\usepackage{pdflscape}  % Landscape pages
\usepackage{multirow}
\usepackage{xcolor}
\usepackage{hyperref}
\usepackage{booktabs}
\usepackage{subfigure}
\usepackage{academicons}
\usepackage{orcidlink}
\usepackage{xspace}

\newcommand{\orcid}[1]{\href{https://orcid.org/#1}{\textcolor[HTML]{A6CE39}{\aiOrcid}}}
\newcommand{\lgal}{\textsc{L-GALAXIES}\xspace}
\newcommand\dd{\mathrm d}
\newcommand\cR{{\mathcal R}_{\rm s}}
\newcommand\manga{\textsc{MaNGA}\xspace}
\newcommand\mangap{\textsc{MaNGA pipe3d}\xspace}

\newcommand{\kms}{\,km\,s$^{-1}$} % kilometres per second
\def\specialname[#1]{\textbf{\textsc{#1}}}

\usepackage[T1]{fontenc}
\usepackage{ae,aecompl}

\usepackage{newtxtext,newtxmath}
\title[]{
  The origin of the galaxy size-stellar metallicity relation - I: A
  semi-analytical perspective
}
\author[Wang et al.]{
  Kai Wang,$^{1, 2}$\thanks{Contact e-mail:
  wkcosmology@gmail.com}\orcidlink{0000-0002-3775-0484}
  \\
  $^{1}$Institute for Computational Cosmology, Department of Physics,
  Durham University, South Road, Durham, DH1 3LE, UK\\
  $^{2}$Centre for Extragalactic Astronomy, Department of Physics,
  Durham University, South Road, Durham DH1 3LE, UK
}

\date{Last updated 2020 May 22; in original form 2018 September 5}

\pubyear{2020}

\begin{document}
\label{firstpage}
\pagerange{\pageref{firstpage}--\pageref{lastpage}}
\maketitle

% Abstract of the paper

\begin{abstract}
  Stellar metallicity encodes the integrated effects of gas inflow, star
  formation, and feedback-driven outflow, yet its connection to galaxy
  structure remains poorly understood. Using SDSS-IV \manga, we present the
  direct observational evidence that, at fixed stellar mass, smaller central
  galaxies are systematically more metal-rich, with a Spearman's rank
  correlation coefficient reaching $\mathcal R_{\rm s}\approx -0.4$. The
  semi-analytical model \lgal reproduces this anti-correlation, albeit with a
  stronger amplitude ($\mathcal R_{\rm s}\approx -0.8$). Within this framework,
  the trend cannot be explained by differences in gravitational potential depth
  or star formation history. Instead, smaller galaxies attain higher stellar
  metallicities because their elevated star formation efficiencies accelerate
  chemical enrichment, and, at fixed stellar mass, they inhabit less massive
  haloes, which makes their recycled inflows more metal-rich. The gas-regulator
  model demonstrates that star formation efficiency affects stellar metallicity
  when the system has not long remained in equilibrium, which is shown to be
  the case for central galaxies with $M_{\rm star}\lesssim 10^{10.5}\rm
  M_\odot$ in both \lgal and observation. The model also suggests a testable
  signature that, at fixed stellar mass, larger or lower-metallicity galaxies
  should inhabit more massive haloes than their smaller and higher-metallicity
  counterparts, providing a direct and testable imprint of the galaxy
  size-stellar metallicity relation on the galaxy-halo connection.
\end{abstract}

% Select between one and six entries from the list of approved keywords.
% Don't make up new ones.
\begin{keywords}
  galaxies: star formation - galaxies: evolution - galaxies: ISM -
  methods: statistical
\end{keywords}

%%%%%%%%%%%%%%%%%%%%%%%%%%%%%%%%%%%%%%%%%%%%%%%%%%

%%%%%%%%%%%%%%%%% BODY OF PAPER %%%%%%%%%%%%%%%%%%

%%%%%%%%%%%%%%%%% SECTION 1 %%%%%%%%%%%%%%%%

\section{Introduction}%
\label{sec:introduction}

The chemical enrichment of galaxies offers a window into the complex interplay
of star formation, feedback, and gas flows that drive their evolution
\citep[][]{maiolinoReMetallicaCosmic2019}. Metals are produced by stars and
redistributed through feedback-driven winds, gas inflows, and recycling
processes, imprinting the integrated history of star formation and gas exchange
with the environment in both the gas and stellar components. This makes
metallicity a cornerstone for understanding how galaxies grow within the cosmic
web, how they regulates their baryons, and how they interact with the host
haloes \citep[e.g.][]{wangEnvironmentalDependenceMassMetallicity2023}.

Much of the progress to date has focused on the gas-phase metallicity. Numerous
surveys have established tight scaling relations between gas metallicity,
stellar mass, and star formation rate, encapsulated in the mass-metallicity and
fundamental metallicity relations in our local Universe
\cite[e.g.][]{tremontiOriginMassMetallicityRelation2004,
  mannucciFundamentalRelationMass2010, zahidUniversalRelationGalactic2014,
maRevisitingFundamentalMetallicity2024} and at high-z Universe
\citep{liMassMetallicityRelationDwarf2023, curtiJADESInsightsLowmass2024,
heEarlyResultsGLASSJWST2024}. These relations are understood as natural
outcomes of gas inflows, outflows, and star formation regulating the
interstellar medium \citep[see][]{finlatorOriginGalaxyMassmetallicity2008,
  lillyGASREGULATIONGALAXIES2013, pengHaloesGalaxiesDynamics2014,
wangGasphaseMetallicityDiagnostic2021}. Variations in galaxy size have also
been linked to gas-phase abundances: at fixed stellar masss, compact galaxies
tend to show higher gas metallicities, a trend that has been attributed to more
efficient retention of metals in deeper potential wells
\citep[e.g.][]{ellisonCluesOriginMassMetallicity2008,
maRevisitingFundamentalMetallicity2024}, or to dilution from recent inflows
preferentially affecting extended galaxies
\citep[e.g.][]{sanchezPipe3DPipelineAnalyze2016}. The galaxy size--metallicity
relation therefore provides a stringent constraint on models of galaxy
formation and evolution.

While gas-phase metallicity provides a snapshot of the present interstellar
medium, stellar metallicities offer a complementary perspective by tracing the
integrated history of galaxies. Large surveys have revealed that stellar
metallicity tightly with stellar mass, giving rise to the stellar
mass-metallicity relation \citep[e.g.][]{gallazziAgesMetallicitiesGalaxies2005,
donnanRoleCosmicWeb2022, garciaInterplayStellarGasphase2024}. Unlike the
gas-phase metallicity, while reflects the ongoing inflows and outflows, stellar
metallicity is less sensitive to short-term fluctuations and instead encodes
the cumulative efficiency with which galaxies have turned gas into stars and
retained their metals \citep{pengStrangulationPrimaryMechanism2015,
luAnalyticalModelGalaxy2015, wangUniversalBimodalityKinematic2024}. Yet,
despite the importance of stellar metallicity, little is known about how it
connects to structural parameters such as galaxy size, even though the
analogous trends are well established for the gas-phase metallicity
\citep{ellisonCluesOriginMassMetallicity2008,
  sanchezalmeidaOriginRelationMetallicity2018,
maRevisitingFundamentalMetallicity2024}.

In this work we combined three complementary approaches to investigate the
connection between galaxy size and stellar metallicity. First, we use \manga
observations of central galaxies to provide the direct evidence for the
galaxy size-stellar metallicity anti-correlation at fixed stellar mass. Second,
we exploit the semi-analytical model \lgal, in which physical processes such as
gas accretion, star formation, and feedback are explicitly prescribed, allowing
us to isolate the potential drivers of this correlation in a controlled
setting. Thirdly, we employ an analytical gas-regulator framework that
clarifies the dependence of stellar metallicity on star formation efficiency
and inflow enrichment, offering physical intuition for the numerical results.
Together, these approaches allow us not only to establish the existence of a
robust anti-correlation between size and stellar metallicity, but also to
identify the physical mechanisms that underpin it.

The remainder of this paper is organized as follows. \S\,\ref{sec:data}
describes the observational and theoretical datasets.
\S\,\ref{sec:size_metallicity_anti_correlation_in_manga} presents the
empirical evidence of the galaxy size-stellar metallicity anti-correlation in
\manga. \S\,\ref{sec:The size-metallicity relation in semi-analytical models}
examines the same relation in \lgal and explores its origin.
\S\,\ref{sec:Connecting stellar metallicity to star formation efficiency}
introduces the gas-regulator framework and connects star formation efficiency
to metal enrichment. \S\,\ref{sec:Discussion} discusses the broader
implications for the galaxy-halo connection and the extension to massive
galaxies, and \S\,\ref{sec:Summary} summarizes our conclusions. Throughout
this paper, we adopted a Planck cosmology
\citep{planckcollaborationPlanck2015Results2016}, in which $h=0.673$,
$\Omega_{\rm m}=0.315$, $\Omega_{\rm b}=0.049$, and $\Omega_{\Lambda}=0.685$.

\section{Data}
\label{sec:data}

\subsection{\manga}
\label{sub:manga} % (fold)

Our observational sample is drawn from the pyPipe3D value-added catalog
\citep{sanchezSDSSIVMaNGAPyPipe3D2022}, which provides spatially resolved
stellar population and ionized-gas properties for galaxies observed by the
SDSS-IV \manga survey \citep{bundyOverviewSDSSIVMaNGA2015}. \manga delivers
optical integral-field spectroscopy for more than 10,000 galaxies, with
wavelength coverage from 3600-10000$\AA$ and a typical spectral resolution of
$R\sim 2000$, spanning the redshift range of $0.01 < z < 0.15$.

The pyPipe3D pipeline \citep{lacerdaPyFIT3DPyPipe3DNew2022} is a Python-based
implementation of Pipe3D \citep{sanchezPipe3DPipelineAnalyze2016,
sanchezPipe3DPipelineAnalyze2016a}, which performs full spectral fitting of
each spaxel in the \manga datacubes using a library of simple stellar population
from the MaStar\_sLOG library \citep{yanSDSSIVMaStarLarge2019}. The library
spans a broad range of stellar ages (1Myr - 13.5Gyr) and metallicities
($Z=0.0001-0.04$). The fitting accounts fro stellar kinematics, internal dust
extinction using the \citet{cardelliRelationshipInfraredOptical1989} law, and
applies adaptive spatial binning to ensure adequate signal-to-noise. This
approach yields robust maps of stellar and gas-phase properties across full
\manga sample, providing the basis for our study.

The total stellar mass, \texttt{log\_Mass}, is obtained by integrating the
dust-corrected stellar mass surface density over the \manga field of view, using
the mass-to-light ratios inferred from the spectral fitting. The mass-weighted
stellar metallicity, \texttt{ZH\_T99}, is the logarithmic mean metallicity of
the stellar population, weighted by stellar mass. The galaxy size,
\texttt{R50\_kpc\_Mass}, is defined as the physical radius enclosing 50\% of
the total stellar mass within the \manga field of view. All quantities are
measured consistently within the spectroscopic field of view, ensuring that
stellar population, structural and integrated properties are physically
matched. To identify central galaxies in the \manga sample, we cross-match these
galaxies with the SDSS group catalog of \citet{yangGalaxyGroupsSDSS2007},
constructed with a halo-based group finder, and select only the most massive
galaxy in each group.

\subsection{L-GALAXIES}
\label{sub:l_galaxies} % (fold)

For theoretical comparison we use central galaxies from the \lgal
semi-analytical model in \citet{ayromlouGalaxyFormationLGALAXIES2021}
\citep[see also][]{springelPopulatingClusterGalaxies2001,
  crotonManyLivesActive2006, deluciaHierarchicalFormationBrightest2007,
  guoDwarfSpheroidalsCD2011, yatesModellingElementAbundances2013,
henriquesGalaxyFormationPlanck2015, henriquesLGALAXIES2020Spatially2020}, built
on halo merger trees from the Millennium simulation
\citep{springelSimulationsFormationEvolution2005} after rescaling to the Planck
cosmology \citep{planckcollaborationPlanck2015Results2016}. In this model, gas
is shock-heated to the virial temperature and cools radiatively onto the
central galaxy, conserving angular momentum and forming an exponential disc
whose initial size is set by the halo spin
\citep[][]{moFormationGalacticDiscs1998}. The disc is resolved into concentric
annuli that track the distribution of cold gas, stars, star formation, and
metals \citep{henriquesGalaxyFormationPlanck2015}. Star formation occurs where
the local gas surface density exceeds a critical threshold
\citep{krumholzStarFormationLaw2009}, and is regulated by stellar feedback,
which ejects gas with a mass-loading that scales with the maximum circular
velocity of haloes; ejected gas is reincorporated on a halo-mass-dependent
timescale. Metals produced by Type II and Ia supernovae and AGB stars are
deposited locally into both gas and stars, providing spatially resolved
predictions of chemical enrichment \citep{yatesModellingElementAbundances2013}.
These prescriptions are implemented explicitly in the semi-analytical
framework, allowing us to identify the impact of individual processes. This
makes \lgal particularly well suited for investigating the connections between
galaxy size, star formation efficiency, and stellar metallicity.

We use all central galaxies at $z=0$ from the public L-GALAXIES catalog of
\citet{ayromlouGalaxyFormationLGALAXIES2021}. To minimise environmental effects
on galaxy size and stellar metallicity, we exclude backsplash systems
\citep[e.g.][]{wangDissectTwohaloGalactic2023}, defined as galaxies that were
once satellites but have subsequently moved beyond the virial radius of their
former hosts. Stellar mass refers to the total mass in both disc and bulge
components, and stellar metallicity is the metal mass fraction normalised by
the solar value $Z_\odot=0.017$. The galaxy size, $r_{\rm star}$, is defined as
the radius enclosing half of the total stellar mass.

\section{Size-metallicity anti-correlation in \manga}
\label{sec:size_metallicity_anti_correlation_in_manga}

\begin{figure}
  \begin{center}
    \includegraphics[width=0.95\linewidth]{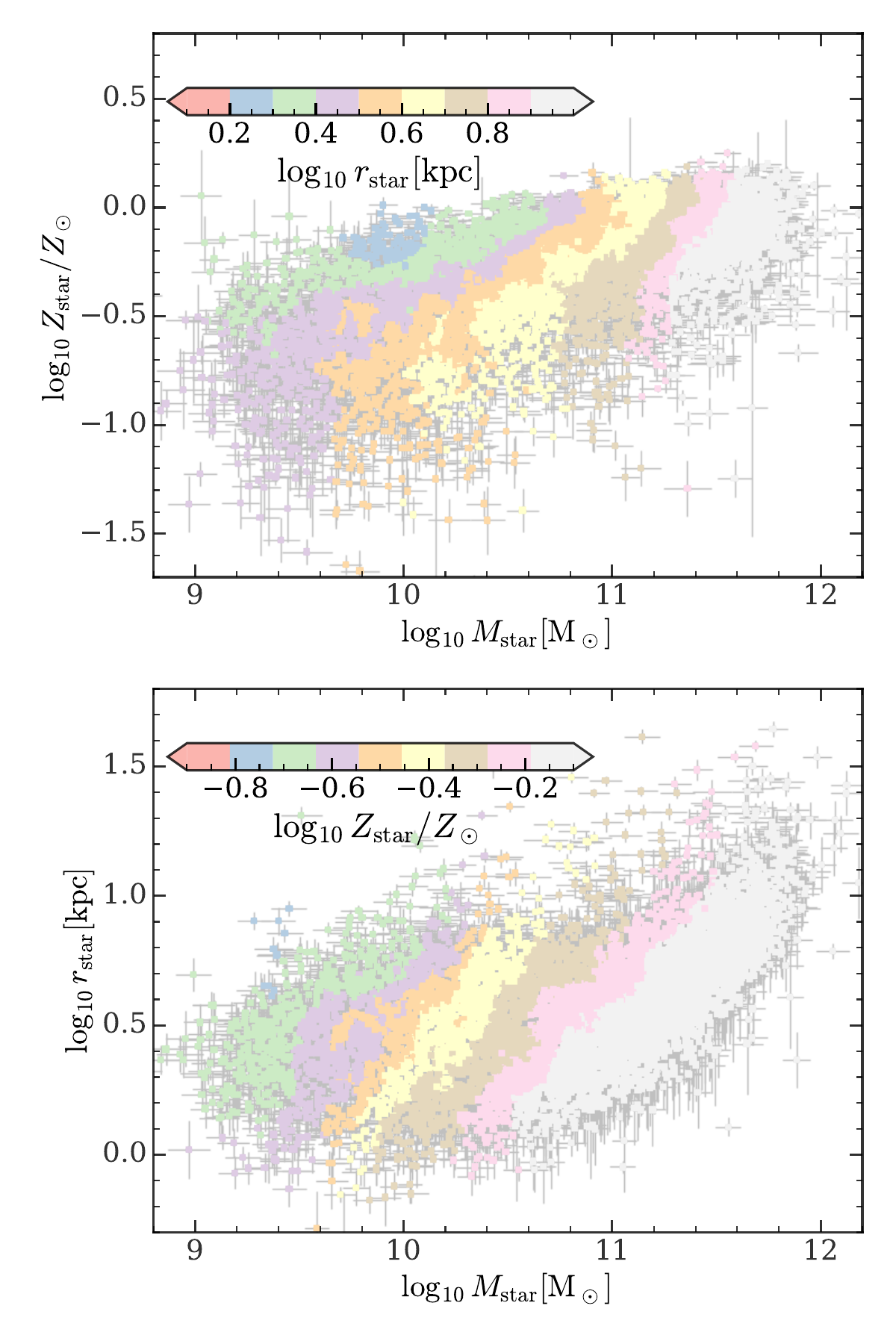}
  \end{center}
  \caption{
    {\bf Top panel:} The stellar mass-stellar metallicity relation with color
    encoding the half-mass size for central galaxies in \mangap. {\bf Bottom
    panel:} The stellar mess-size relation with color encoding the stellar
    metallicity for the same galaxy sample. At fixed stellar mass, smaller
    central galaxies have higher stellar metallicity than their extended
    counterparts. Similarly, higher-$Z_{\rm star}$ central galaxies are smaller
    than their lower-$Z_{\rm star}$ counterparts at fixed stellar mass.
  }
  \label{fig:manga_distribution}
\end{figure}

\begin{figure*}
  \begin{center}
    \includegraphics[width=0.8\linewidth]{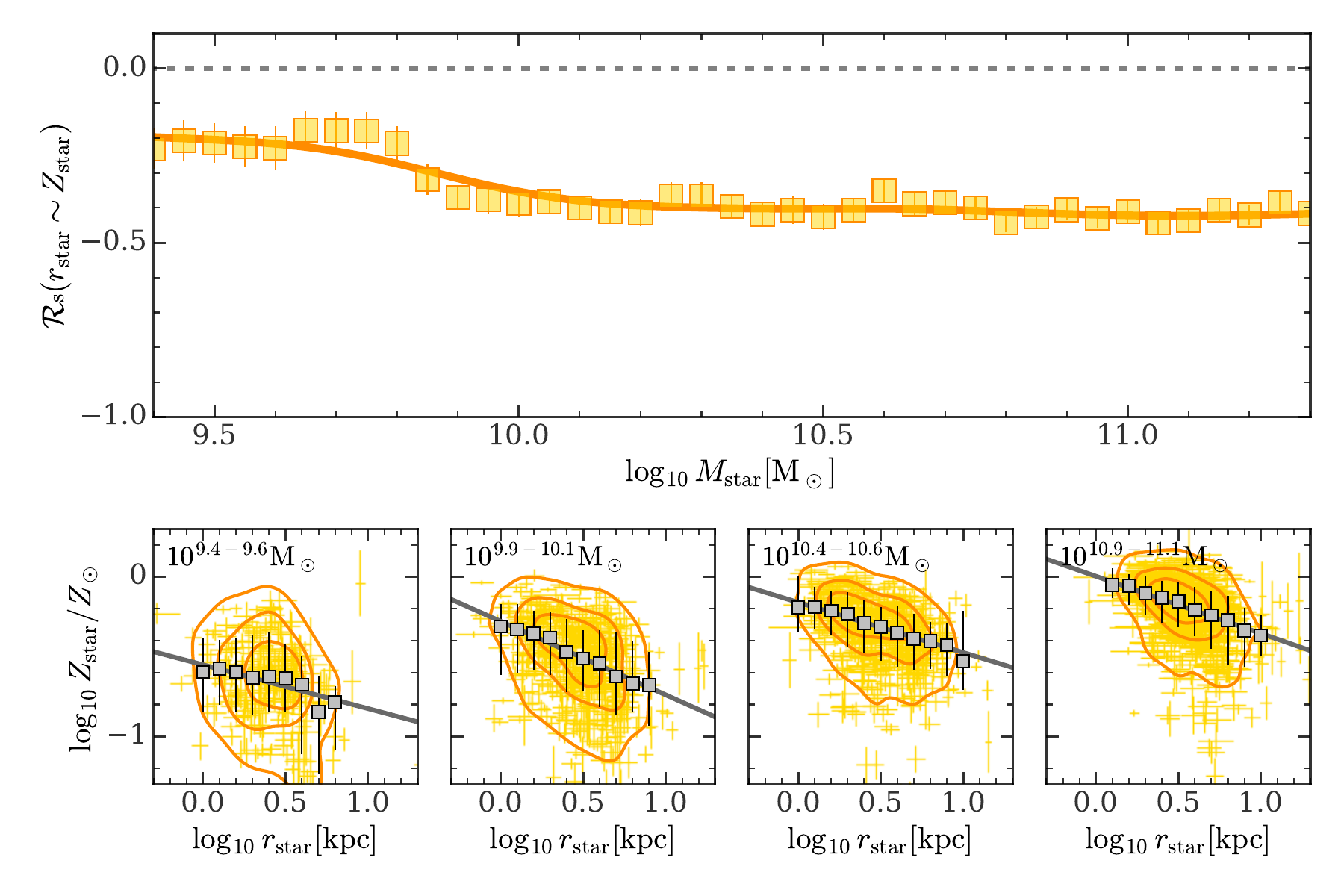}
  \end{center}
  \caption{
    {\bf Top panel:} Spearman's rank correlation coefficients between galaxy
    half-mass size and the stellar metallicity in 0.2-dex stellar mass bins for
    central galaxies in \mangap, with error bars show the standard deviation of
    100 bootstrap sample. The solid line is a smoothing B-spline fit to the
    data points to show the trend. {\bf Bottom panels:} The joint distribution
    of stellar metallicity and galaxy half-mass radius in selected stellar mass
    bins. The gray boxes shows the median stellar metallicity in bins of galaxy
    sizes, with the error bar shows 16-84th percentiles. The grey solid line
    shows the linear fitting to the median trend. In fixed stellar mass bins,
    stellar metallicity and galaxies are anti-correlated to each other, and the
    strength of this correlation increases from $\approx -0.2$ at $M_{\rm
    star}\sim 10^{9.5}\rm M_\odot$ to $\approx -0.4$ above $10^{10}\rm
    M_\odot$.
  }
  \label{fig:manga_correlation}
\end{figure*}

With the dataset defined, we now examine the relation between galaxy size and
stellar metallicity in \manga. This provides the direct observational
evidence for a galaxy size-stellar metallicity anti-correlation at fixed
stellar mass. Fig.~\ref{fig:manga_distribution} shows the stellar
mass-metallicity and stellar mass-size joint distributions for central galaxies
in \manga, color-coded by galaxy size and stellar metallicity, respectively. At
fixed stellar mass, smaller galaxies are systematically more metal-rich. This
trend is evident across the entire stellar mass range.

To quantify the relation between size and metallicity, we compute Spearman's
rank correlation coefficients in 0.2-dex stellar mass bins, requiring at least
30 galaxies per bin, in Fig.~\ref{fig:manga_correlation}. The correlation is
weak at $M_{\rm star}\sim 10^{9.5}\rm M_\odot$ ($\cR\approx -0.2$), but
strengthens towards higher masses, reaching $\cR\approx -0.4$ above $\approx
10^{10}\rm M_\odot$. The lower panels show the joint distribution of size and
metallicity, with contours enclosing 10, 40, and 70 per cent of the sample.
Median metallicities in 0.2-dex bins (grey boxes) confirm a clear
anti-correlation between these two quantities.

This study presents the direct observational evidence for an anti-correlation
between galaxy size and stellar metallicity \citep[see
also][]{vaughanSAMIGalaxySurvey2022, boardmanCompetingEffectsRecent2025,
liCentralVelocityDispersion2025}, while analogous trends between galaxy size
and gas-phase metallicity have been identified in numerous studies. For
example, \citet{ellisonCluesOriginMassMetallicity2008} demonstrated that, at
fixed stellar mass, galaxies with larger half-light radii tend to have lower
gas-phase metallicity by up to $\sim 0.2$ dex. A range of explanations have
been proposed for this relation. One view is that compact galaxies are able to
retain metals more effectively because their deeper gravitational potentials
suppress the efficiency of feedback-driven outflows
\citep[e.g.][]{ellisonCluesOriginMassMetallicity2008,
maRevisitingFundamentalMetallicity2024}. Another interpretation links the trend
to recent gas accretion: the inflow of metal-poor material both dilutes the
interstellar medium and contributes to the growth of galaxy outskirts,
producing lower metallicities in larger systems
\citep[e.g.][]{sanchezalmeidaOriginRelationMetallicity2018}. A third
explanation emphasises the star formation efficiency (SFE), defined as the
ratio between star formation rate and cold gas mass, since galaxies with
smaller radii at fixed stellar mass have higher surface densities and shorter
depletion times, leading to more efficient enrichment of their gas reservoirs
\citep[e.g.][]{ellisonCluesOriginMassMetallicity2008,
sanchezalmeidaOriginRelationMetallicity2018}.

While these mechanisms have been discussed extensively in the context of
gas-phase metallicities, the extension to the stellar component is less
straightforward: gas-phase metallicity reflects the present-day chemical state
of the interstellar medium, whereas stellar metallicity records the cumulative
enrichment history of a galaxy. Observations further indicates that the offset
between the two tracers is not fixed but depends on galaxy properties
\citep{fraser-mckelvieSAMIGalaxySurvey2022,
zinchenkoDifferentBehaviourGasphase2024}. These differences underscore the need
to treat stellar metallicity as a distinct diagnostic, and motivate the use of
models that can capture long-term evolutionary processes.

\section{The size-metallicity relation in semi-analytical models}
\label{sec:The size-metallicity relation in semi-analytical models}

\begin{figure*}
  \begin{center}
    \includegraphics[width=0.8\linewidth]{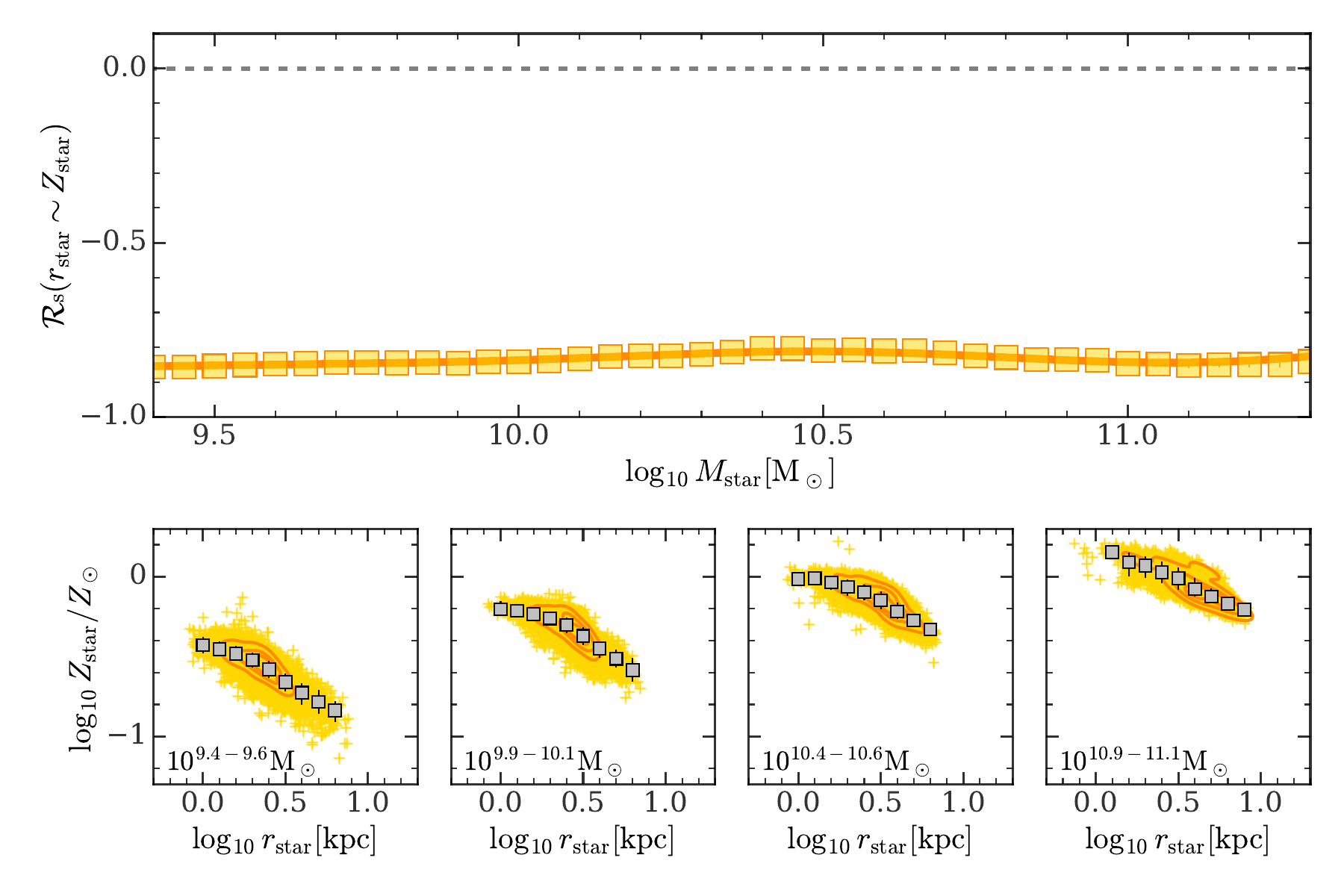}
  \end{center}
  \caption{ {\bf Top panel:} Spearman's correlation coefficients galaxy
    half-mass stellar size and stellar metallicity for central galaxies in
    \lgal calculated in 0.1-dex-width stellar mass bins. {\bf Bottom panels:}
    The joint distribution of galaxy size and stellar metallicity in four
    selected stellar mass bins. The contour lines enclose 10, 40, and 70\% of
    galaxies in each subsample. The grey boxes show the median stellar
    metallicity as a function of galaxy size, with error bars show the 16-84th
    percentiles. \lgal shows a strong anti-correlation between stellar
    metallicity and galaxy size, and the correlation strength is $\approx -0.8$
    across the whole stellar mass range.
  }
  \label{fig:lgal_correlation}
\end{figure*}

Understanding the origin of the galaxy size-stellar metallicity relation
requires a model that connects galaxy structure to the cumulative effects of
star formation and enrichment. Semi-analytical models are well suited for this
purpose, due to that all physical processes are explicitly prescribed,  which
makes it easy to isolate the impact of individual processes. Even though such
models are simplified by construction, they provide a useful framework to test
whether the observed trend arises naturally from current prescriptions and to
identify the mechanisms that are most likely responsible. In particular, the
\lgal model, which reproduces a wide range of observed galaxy scaling
relations \citep[e.g.][]{yatesRelationMetallicityStellar2012}, offers an ideal
laboratory for investigating the physical origin of the galaxy size-stellar
metallicity relation.

Fig.~\ref{fig:lgal_correlation} shows the correlation between galaxy size and
stellar metallicity in \lgal, analysed in the same stellar mass bins as for
\manga. A strong anti-correlation is evident, with a Spearman's correlation
coefficient of $\mathcal R_{\rm s}\approx -0.8$ across the full stellar mass
range, which is significantly stronger than observed in \manga. The lower
panels display the joint distribution, with grey boxes marking the median
stellar metallicity as a function of size. The relation is relatively shallow
below $r_{\rm star}\lesssim 10^{-0.3}\rm kpc$ ($\approx 2\rm kpc$), with a
slope of $\approx -0.3$, but steepens to $\approx -0.6$ at larger radii.

Comparing to the observational results in Fig.~\ref{fig:manga_correlation}, we
find that \lgal reproduces the galaxy size-stellar metallicity relation of
compact galaxies ($r_{\rm star}\sim 1\rm kpc$) remarkably well. For more
extended galaxies, however, \lgal predicts stellar metallicities that are
$\approx 0.05-0.2$ dex lower than observed, leading to a stronger overall
anti-correlation between galaxy size and stellar metallicity. This close
agreement motivates a closer examination of the physical origin of the relation
within \lgal.

\subsection{Roles of gravitational potential and star formation history}
\label{sub:Roles of potential depth and star formation history} % (fold)

\begin{figure}
  \begin{center}
    \includegraphics[width=0.95\linewidth]{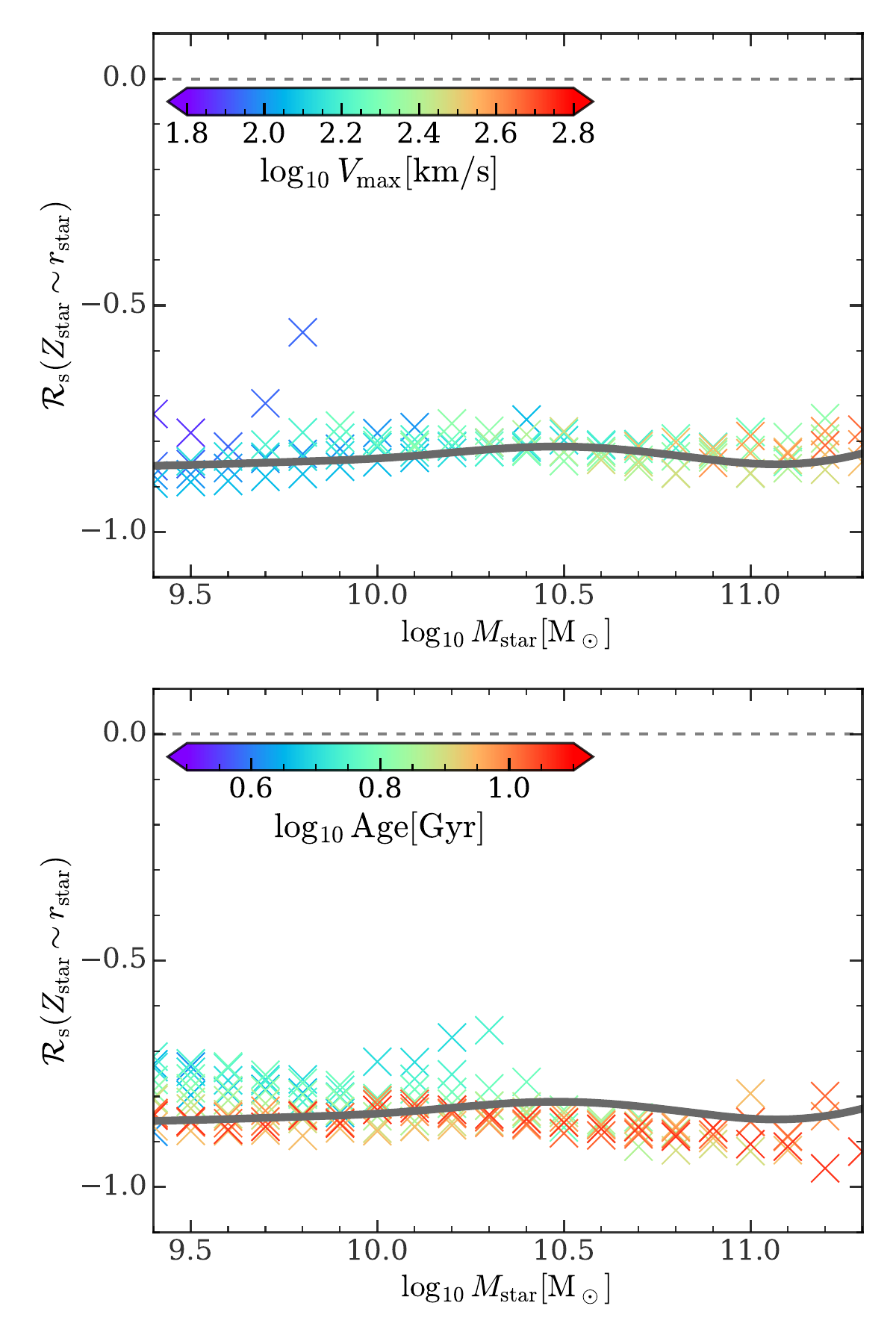}
  \end{center}
  \caption{
    {\bf Top panel:} The symbols are Spearman's rank correlation coefficients
    between galaxy stellar size and galaxy stellar metallicity with in bins of
    stellar mass (0.1 dex) and $V_{\rm max}$ (0.1 dex). {\bf Bottom panel:} The
    symbols are Spearman's rank correlation coefficients between galaxy stellar
    size and galaxy stellar metallicity with in bins of stellar mass (0.1 dex)
    and stellar age (0.1 dex). In both panels, the gray solid line shows the
    rank correlation coefficients with only stellar mass fixed. The
    anti-correlation between galaxy stellar size and galaxy stellar metallicity
    remains at the same level after fixing $V_{\rm max}$. This indicates that
    either gravitational potential or star formation history is responsible for
    the anti-correlation between galaxy stellar size and stellar metallicity.
  }
  \label{fig:lgal_vmax_age}
\end{figure}

Previous studies of gas-phase metallicity have suggested that galaxy
metallicity traces the depth of the gravitational potential: compact galaxies
are expected to retain metals more efficiently because their deeper
gravitational potential, contributed by both concentrated baryons and
contracted dark matter distribution, suppress feedback-driven outflow
\citep[e.g.][]{lillyGASREGULATIONGALAXIES2013, pengHaloesGalaxiesDynamics2014,
maRevisitingFundamentalMetallicity2024}. A key advantage of semi-analytical
models is that the relevant processes are explicitly prescribed, allowing us to
control the effect of individual prescriptions. In \lgal, stellar feedback
depends only on the maximum circular velocity of the host halo, $V_{\rm max}$.
Therefore we can fix this property and then investigate the correlation between
galaxy size and stellar metallicity.

The top panel of Fig.~\ref{fig:lgal_vmax_age} shows the correlation between
galaxy size and stellar metallicity with both stellar mass and $V_{\rm max}$
fixed. Spearman's rank correlation analysis shows that they are still strongly
correlated to each other with coefficient values of $\approx -0.8$ when $V_{\rm
max}$ is fixed. These results suggest that, within \lgal, gravitational
potential depth alone does not account for the galaxy size-stellar metallicity
correlation.

Another proposed explanation for the correlation between galaxy size and
gas-phase metallicity is that it reflects differences in star formation
history. For example, Recent accretion of metal-poor gas both dilutes the
interstellar medium and builds up the outskirts of galaxies, potentially
producing lower gas-phase metallicities in more extended galaxies
\citep[e.g.][]{sanchezalmeidaOriginRelationMetallicity2018}. To test this
scenario in \lgal, we use mass-weighted stellar age as a proxy for star
formation history and compute Spearman's correlation coefficients between
galaxy size and stellar metallicity in finer bins of stellar age, as shown in
the bottom panel of Fig.~\ref{fig:lgal_correlation}. Their correlation
coefficients are close to each other for galaxies with different stellar ages.
This outcome is not unexpected: while recent inflows can strongly lower
gas-phase metallicities, they leave the stellar metallicity largely unaffected
because it reflects the cumulative enrichment of past star formation. We
therefore conclude that variations in star formation history do not drive the
size-metallicity relation in \lgal.

In addition to the conditional correlation analysis,
Appendix~\ref{sec:galaxy_size_stellar_metallicity_relation_in_lgal} shows the
galaxy size-stellar metallicity joint distribution in fine bins of $V_{\rm
max}$ and stellar age, also with stellar mass fixed. One can see that
not only the strength of anti-correlation, but also the median stellar
metallicity as a function of galaxy size, is not affected by changing $V_{\rm
max}$, nor stellar age.

Taken together, these tests show that, in \lgal, neither gravitational
potential depth nor variations in star formation history drive the
size-metallicity relation. This outcome reflects a deliberate feature of the
semi-analytical framework: the mass-loading factor is prescribed explicitly as
a function of $V_{\rm max}$, which allows us to isolate and control its
influence on star-formation efficiency. This level of direct manipulation is
not straightforward in hydrodynamical simulations, where the outflow efficiency
emerges from complex gas dynamics and subgrid implementations, and cannot be
tuned independently of other processes. The model therefore enables us to
identify star formation efficiency as a substantial contributor to the
size-metallicity correlation. We emphasise that this does not diminish the
physical relevance of the result; rather, it clarifies which processes are
accessible in this modelling framework. Guided by this insight, future work can
test the relative roles of potential depth and star formation efficiency in
hydrodynamical simulations in a more targeted and physically interpretable
manner.

\subsection{Star formation efficiency in \lgal}
\label{sub:Star formation efficiency in lgal} % (fold)

\begin{figure*}
  \begin{center}
    \includegraphics[width=0.95\linewidth]{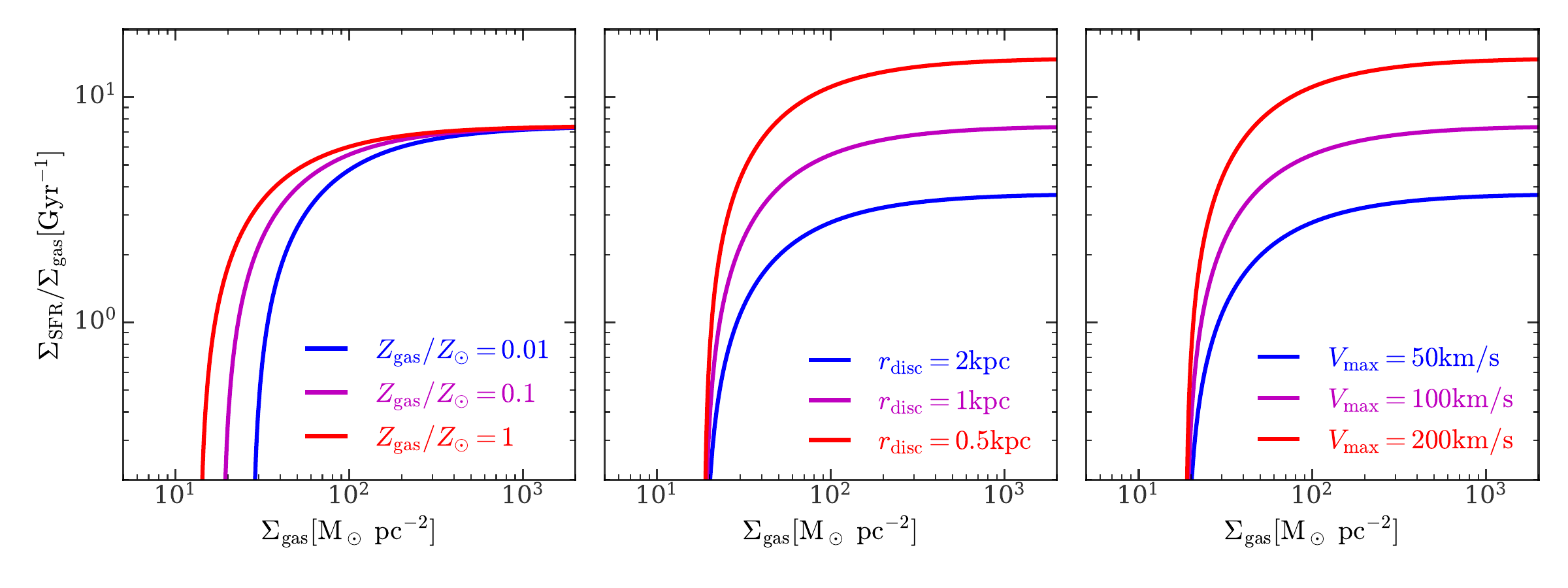}
  \end{center}
  \caption{
    The surface SFE, $\Sigma_{\rm SFR}/\Sigma_{\rm gas}$ as a function of gas
    surface density $\Sigma_{\rm gas}$, for the \lgal implementation in
    \citet{ayromlouGalaxyFormationLGALAXIES2021}. Results are shown for
    different gas metallicities ($Z_{\rm gas}$, left panel), galaxy sizes
    ($r_{\rm disc}$, middle panel), and maximum circular velocities of the host
    halo ($V_{\rm max}$, right panel). The surface SFE is
    regulated by these three factors, in a sense that higher $Z_{\rm gas}$,
    smaller $r_{\rm disc}$, and higher $V_{\rm max}$ lead to higher SFE.
  }
  \label{fig:lgal_sfe}
\end{figure*}

Having ruled out gravitational potential depth and star formation history, we
now examine the role of SFE in \lgal. Star formation is computed in concentric
annuli, where the star formation rate surface density is
\begin{equation}
  \Sigma_{\rm SFR} = \alpha_{\rm H_2}f_{\rm H_2}\Sigma_{\rm gas}/t_{\rm dyn}
\end{equation}
where $t_{\rm dyn} = R_{\rm cold, disc}/V_{\rm max}$, $\Sigma_{\rm gas}$ is the
cold gas surface density, $\alpha_{\rm H_2}$ is the model parameter. The
molecular fraction, $f_{\rm H_2}$, follows the
\citet{krumholzStarFormationLaw2009} prescription, which depends on
$\Sigma_{\rm gas}$ and the gas-phase metallicity. The cold gas disc
radius $R_{\rm cold, disc}$ is set by the halo spin, linking compactness
directly to SFE.

Fig.~\ref{fig:lgal_sfe} shows the resulting SFE, $\Sigma_{\rm SFR}/\Sigma_{\rm
gas}$, as a function of gas surface density. The SFE rises with gas surface
density until $\Sigma_{\rm gas}\gtrsim 100\rm M_\odot\,pc^{-2}$, where the
molecular fraction saturates and SFE flattens. At low densities ($\approx 20\rm
M_\odot\,pc^{-2}$), star formation is strongly suppressed, with a threshold
that depends moderately on gas metallicity. Galaxies with smaller sizes and
higher $V_{\rm amx}$ attain higher SFE, reflecting both higher surface
densities and shorter dynamical times.

We postulate that, at fixed stellar mass, forming a small disc galaxy requires
a compact gaseous disc, since the stellar disc inherits the angular momentum
from the gaseous disc \citep{moFormationGalacticDiscs1998}. Compact gaseous
discs have higher gas surface density and shorter dynamical timescale, which
combined together enhance the SFE significantly
\citep{leroyStarFormationEfficiency2008}. Finally, a higher SFE enrich the
interstellar medium faster so that there are more metals locked in to the
stellar population.

It is important to note, however, that in the equilibrium state
\citep[][]{lillyGASREGULATIONGALAXIES2013, pengHaloesGalaxiesDynamics2014},
stellar metallicity is set by the mass loading factor and independent of SFE.
This implies that SFE can only influence the stellar metallicity when galaxies
are out of equilibrium. We will demonstrate it with a gas-regulator model in
the next section.

\section{Connecting stellar metallicity to star formation efficiency}
\label{sec:Connecting stellar metallicity to star formation efficiency}

\begin{figure}
  \begin{center}
    \includegraphics[width=0.95\linewidth]{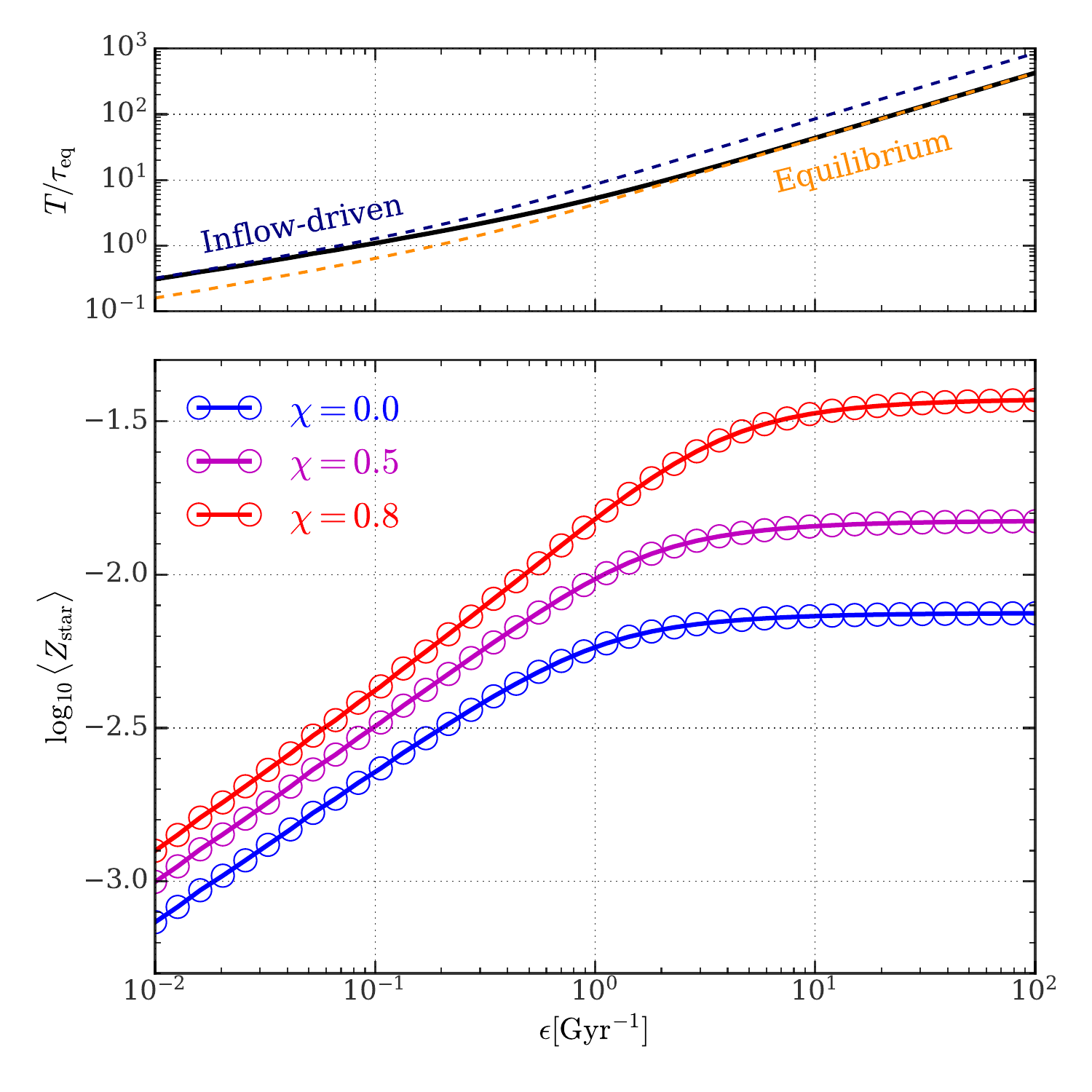}
  \end{center}
  \caption{
    {\bf Top panel:} The ratio between galaxy evolution timescale, $T$, and the
    equilibrium timescale, $\tau_{\rm eq}$, as a function of star formation
    efficiency, $\epsilon$. The dark blue and orange dashed line show the
    analytical result in the inflow-driven and equilibrium limits,
    respectively. {\bf Bottom panel:} The stellar metallicity, $\langle Z_{\rm
    star}\rangle$, of galaxies at targeted stellar mass, $M_{\rm star, T}$, as
    a function of star formation efficiency within different inflow pollution
    parameter, $\chi$. Stellar metallicity increases with star formation
    efficiency when $T/\tau_{\rm reg}\ll 20$, after which stellar metallicity
    saturates and exhibits minimal dependence on SFE. Stellar metallicity is
    also subject to the metallicity of inflow gas, which is quantified by
    $\chi$, and the dependence is stronger in the equilibrium limit than in the
    inflow-driven limit.
  }
  \label{fig:model}
\end{figure}

We have shown that galaxy size regulates SFE in \lgal. The remaining question
is whether, and to what extent, SFE shapes the cumulative stellar metallicity.
To address this, we turn to the gas-regulator framework, which describes the
balance of gas inflow, star formation, feedback-driven outflows, and the
recycling of enriched material. This model provides a simple yet physically
transparent link between SFE and chemical enrichment, and allows us to isolate
the role of SFE in setting stellar metallicities.

\subsection{Equations governing the evolution}
\label{sub:Equations governing the evolution} % (fold)

The gas-regulator model is governed by coupled continuity equations for the gas
mass and its metal content:
\begin{align}
  \frac{\dd M_{\rm gas}}{\dd t} &= \Phi - (1-R + \eta)\Psi \label{eq:mgas}\\
  \frac{\dd (M_{\rm gas}Z_{\rm gas})}{\dd t}&= y\Psi - Z_{\rm
  gas}(1-R + \eta)\Psi + Z_{\rm in}\Phi
  \label{eq:zgas}
\end{align}
where $\Phi$ is the inflow rate, $\Psi$ the star formation rate, $R$ the
return fraction, $\eta$ the mass-loading factor, and $y$ the yield. We assume
a constant SFE $\epsilon=\Psi / M_{\rm gas}$, and for
simplicity a constant inflow rate $\Phi$. The inflowing gas may be pristine
or enriched by recycled outflows; we quantify this with $\chi\equiv Z_{\rm
in}/Z_{\rm gas}$, which ranges from $\chi=0$ (pristine inflow) to 1
(reaccreted material with the same metallicity as the interstellar medium).

Solving equations~\ref{eq:mgas}-\ref{eq:zgas} yields the gas mass, $M_{\rm
gas}(t)$, and gas metallicity, $Z_{\rm gas}(t)$. The corresponding stellar mass
and mass-weighted stellar metallicity follow from their standard integral
definitions,
\begin{align}
  M_{\rm star}(t) &\equiv \int_0^t(1-R)\epsilon M_{\rm
  gas}(t^\prime)\dd t^\prime \\
  \langle Z_{\rm star}\rangle(t) &\equiv \frac{1}{M_{\rm
  star}(t)}\int_0^t(1-R)\epsilon M_{\rm gas} (t^\prime) Z_{\rm
  gas}(t^\prime) \dd t^{\prime}
\end{align}
The evolution is controlled by two characteristic timescales. The first
is the equilibrium timescale,
\begin{equation}
  \tau_{\rm eq} = \frac{1}{(1-R + \eta)\epsilon}
\end{equation}
which sets how rapidly gas approaches a quasi-steady state \citep[see
also][]{pengHaloesGalaxiesDynamics2014}. The second is the effective
growth timescale, $T$, defined as the time required to build up a target
stellar mass $M_{\rm star, T}$, under constant inflow. $T$ need not equal
the total age of the system, as it only counts periods of active star
formation supplied by the constant inflow, and is typically shorter for
galaxies with bursty histories.

\subsection{Inflow-driven limit}
\label{sub:Inflow-driven limit} % (fold)

Consider the case where the effective growth timescale is much shorter than the
equilibrium timescale, $T\ll \tau_{\rm eq}$. In this regime the system does not
reach equilibrium, and the gas mass evolve as
\begin{equation}
  M_{\rm gas}(t) = \Phi\tau_{\rm eq} + \left(M_{\rm gas, 0} -
  \Phi\tau_{\rm eq}\right)\mathrm e^{-t / \tau_{\rm eq}}
  \stackrel{t\ll \tau_{\rm eq}}{\approx} \Phi t
\end{equation}
where we have neglected the initial gas reservoir. The gas content is therefore
set directly by the inflow, motivating the term ``inflow-driven limit''.

The corresponding stellar mass evolution is
\begin{equation}
  M_{\rm star, T} = \int_0^{T}(1-R)\epsilon M_{\rm gas}(t^\prime)\dd
  t^\prime = \frac12 (1-R)\epsilon\Phi T^2
  \label{eq:mstar_t}
\end{equation}
And the gas-phase metallicity is
\begin{equation}
  Z_{\rm gas}(t) = \frac{\epsilon y}{2-\chi}t =
  \frac{y}{2-\chi}\sqrt{\frac{2\epsilon M_{\rm star, T}}{(1-R)\Phi}}
\end{equation}
Combining these results, the stellar metallicity at $t=T$ is
\begin{equation}
  \langle Z_{\rm star} \rangle = \frac23 Z_{\rm gas}(T) = \frac23
  \frac{y}{2-\chi}\sqrt{\frac{2\epsilon M_{\rm star, T}}{(1-R)\Phi}}
\end{equation}
Thus, in the inflow-driven regime the stellar metallicity increases with SFE: a
higher SFE accelerates enrichment and locks more metals into stars before the
system reaches equilibrium. The level of enrichment also depends on the
composition of the inflowing gas, scaling with $1 / (2 - \chi)$, so that
reaccreted, metal-rich inflows boost stellar metallicities relative to pristine
accretion.

\subsection{Equilibrium limit}
\label{sub:Equilibrium limit} % (fold)

In the opposite regime, where the evolutionary timescale is much longer
than the equilibrium timescale ($\tau_{\rm eq}$), the system settles
into a quasi-steady state in which both the gas mass and gas metallicity
are approximately time-independent. Strictly speaking, the earliest phase
of non-equilibrium evolution is imprinted in the stellar population, but
once the equilibrium phase dominates the stellar mass budget, the equilibrium
solution describes the bulk properties.

Setting $\dd M_{\rm gas}/\dd t = 0$ and $\dd Z_{\rm gas}/\dd t=0$ yields the
equilibrium gas metallicity:
\begin{equation}
  Z_{\rm gas} = \frac{y}{1-R + \eta}\frac{1}{1-\chi}
\end{equation}
Over sufficiently long times, the stellar metallicity converges to this value
$\langle Z_{\rm star}\rangle \approx Z_{\rm gas}$. Besides, the gas mass is
maintained at a constant level $M_{\rm gas} = \Phi \tau_{\rm eq}$, and the
stellar mass keeps growing as $M_{\rm star}(t) = (1 - R)\epsilon \tau_{\rm
eq}\Phi t$.

In this limit the stellar metallicity becomes independent of SFE. Instead, it
is regulated primarily by the metal content of the inflow, parameterized by
$\chi$. If accreting gas is predominately pristine ($\chi \to 0$), stellar
metallicities reflect the simple balance of yields and outflows. If, however,
reaccreted enriched material dominates ($\chi\to 1$), the equilibrium
metallicity can be substantially boosted.

The regulator model therefore highlights a fundamental distinction between the
inflow-driven and equilibrium regimes: in the former, SFE accelerates
enrichment while regulated by the inflow gas metallicity, whereas in the
latter, SFE drops out and the enrichment level is only set subject to the
inflow metallicity.

\subsection{Numerical solution}
\label{sub:Numerical solution} % (fold)

To explore the full dependence of stellar metallicity on SFE, we numerically
integrate equations~\ref{eq:mgas}-\ref{eq:zgas} until the stellar mass reaches
a fixed target value, which is set to $M_{\rm
star}=10^{10}\rm M_\odot$ here. The fiducial parameters are $\Phi=10^{10}\rm
M_\odot/Gyr$, $\eta = 1$, $R=0.4$, and $y=0.02$, with SFE varies over
$0.01\leq\epsilon \leq 100\rm Gyr^{-1}$. We consider three values of
$\chi\in [0,
0.5, 0.8]$ to represent different levels of inflow enrichment.

The results are shown in Fig.~\ref{fig:model}. Three features stand out.
Firstly, the ratio $T/\tau_{\rm eq}$ between the growth and equilibrium
timescales increases monotonically with $\epsilon$, mapping directly onto the
transition between inflow-driven and equilibrium regimes. Secondly, stellar
metallicity increases systematically with $\chi$, as reaccreted enriched
inflows raise the equilibrium level; the sensitivity to $\chi$ is weak
($\propto 1 / (2-\chi)$) when $T/\tau_{\rm eq}\ll 1$ and strong ($\propto 1 /
(1 -\chi)$) when $T/\tau_{\rm eq}\gg 1$. Thirdly, at fixed $\chi$, the stellar
metallicity initially grows with SFE as $\langle Z_{\rm star}\rangle\propto
\sqrt{\epsilon}$, but once $T/\tau_{\rm eq}\gtrsim 20$, the growth of stellar
metallicity saturates and it becomes independent of SFE. These behaviours
mirror the analytical limits derived above: an SFE-driven scaling in the
inflow-driven regime and an SFE-independent plateau in equilibrium.

\subsection{Evolution stages of galaxy system}
\label{sub:evolution_stages_of_galaxy_system} % (fold)

\begin{figure}
  \begin{center}
    \includegraphics[width=0.95\linewidth]{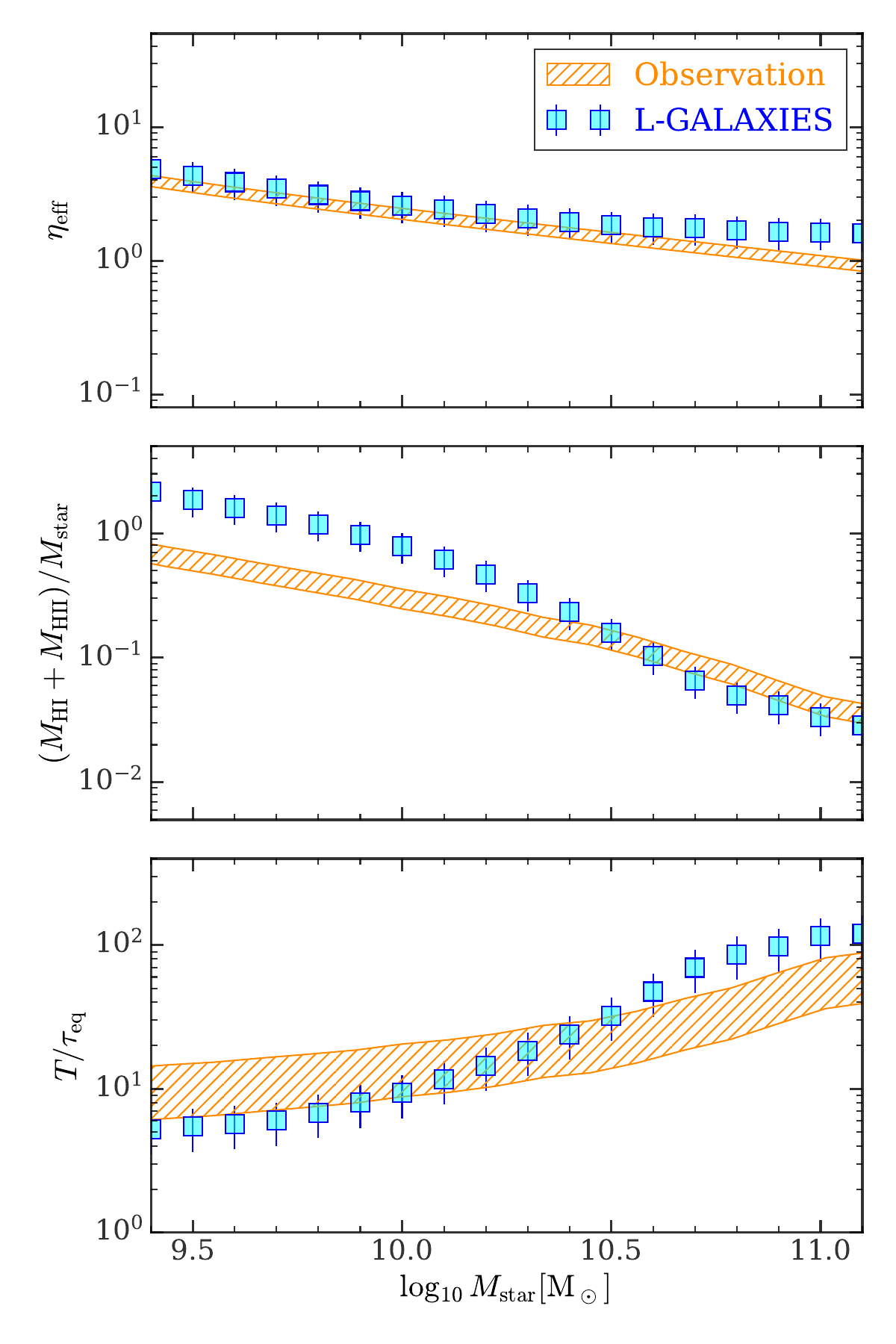}
  \end{center}
  \caption{
    {\bf Top panel:} The effective loading factor, $\eta_{\rm eff}$, estimated
    using the method in \citet{luAnalyticalModelGalaxy2015} as a function of
    stellar mass. The observation result is taken from
    \citet{linConstraintsGalacticOutflows2023}. {\bf Middle panel:} The
    gas-to-stellar mass ratio as a function of stellar mass. The observational
    result is taken from \citet{guoNeutralUniverseMachineEmpiricalModel2023}.
    {\bf Bottom panel:} The timescale ratio, $T/\tau_{\rm eq}$, as a function
    of stellar mass estimated using
    equations~\ref{eq:ratio_1}-\ref{eq:ratio_2}. Our estimation shows that
    galaxies below $10^{10.5}\rm M_\odot$ have $T/\tau_{\rm eq} \lesssim 20$,
    making their stellar metallicity still sensitive to the variation of SFE.
  }
  \label{fig:ratio_estimate}
\end{figure}

Having established the connection between SFE and stellar metallicity, we still
need to show that the galaxy systems we are studying in both observation and
\lgal are not far away from the inflow-driven limit, which means $T/\tau
\lesssim 20$, for SFE to affect the stellar metallicity.

This timescale ratio, $T/\tau$, essentially reflects the competition between
gas inflow and gas consumption due to star formation and outflow, and their
relative dominance over the gas content in galaxies. If the gas content is
dominated by the inflow process, i.e. $M_{\rm gas}\approx \Phi T$, then the
system is in the inflow-driven limit. On the other side, if the star formation,
or feedback, process is sufficiently efficient to balance the inflow process,
the total gas mass would be maintained at a constant level, i.e. $M_{\rm
gas}\approx \Phi\tau_{\rm eq}$, and the system is in equilibrium. Meanwhile, we
notice that the timescale ratio $T/\tau$ is independent of the inflow rate
$\Phi$. This is because that the inflow rate only affects how fast the galaxy
system grows, so that a higher inflow rate gives higher star formation rate and
higher outflow rate, which is unrelated to the evolution stage of the galaxy.
Therefore, we should be able to estimate the timescale ratio from the current
status of the galaxy system.

Indeed, we find that, in the inflow-driven limit, we have
\begin{equation}
  \left.\frac{T}{\tau_{\rm eq}}\right\vert_{\rm inflow-driven} = \frac{2(1 - R
  + \eta)}{(1 - R)\mu}
  \label{eq:ratio_1}
\end{equation}
where $\mu \equiv M_{\rm gas}/M_{\rm star, T}$ is the gas-to-stellar mass
ratio. Similarly, the equilibrium limit gives
\begin{equation}
  \left.\frac{T}{\tau_{\rm eq}}\right\vert_{\rm equilibrium} = \frac{(1 - R +
  \eta)}{(1 - R)\mu}
  \label{eq:ratio_2}
\end{equation}
The results for these two limit cases differ only by a factor of two, thus
they combined together give us a rather accurate estimate of $T/\tau$ of
individual galaxy system.

To apply to galaxies in \lgal and observation, we still need to estimate the
loading factor $\eta$. This is challenging since $\eta$ is subject to the
gravitational potential depth of the galaxy, and the potential is evolving as
the galaxy grows. Thus, we choose to use the loading factor averaged over the
whole evolution history, denoted as $\eta_{\rm eff}$, as a proxy of this
parameter, which can be estimated using the recipe introduced in
\citet{luAnalyticalModelGalaxy2015} (see
  Appendix~\ref{sec:estimating_the_effective_loading_factor} for a detailed
derivation). It should be noted that this estimate converges to the true value
in a gas-regulator model in which the loading factor is set to a constant,
regardless whether the system is in the inflow-driven limit or in equilibrium.

The top panel of Fig.~\ref{fig:ratio_estimate} shows the estimated effective
loading factor $\eta_{\rm loadin}$ as a function of stellar mass for all
central galaxies in \lgal. Combined with the gas-to-stellar mass ratio in the
middle panel, we can estimate the ratio $T/\tau_{\rm eq}$ and show the result
in the bottom panel. Galaxy systems have just entered the equilibrium state
($T/\tau_{\rm eq} \lesssim 20$) for galaxies with $M_{\rm star}\lesssim
10^{10.5}\rm M_\odot$, and the ratio increases with stellar mass. However, as
we will discuss in \S\,\ref{sub:massive}, these massive galaxies are assembled
through {\it ex-situ} mergers rather than {\it in-situ} star formation, so the
gas-regulator model does not apply here, and the anti-correlation between
galaxy size and stellar metallicity is maintained through the merging process.

We also estimate $T/\tau_{\rm eq}$ for observed galaxies in our local Universe
using empirical scaling relations in bins of stellar mass. As shown in
equations~\ref{eq:ratio_1} and \ref{eq:ratio_2}, this requires the mass loading
factor and the gas mass fraction. We adopt the observationally inferred loading
factor from \citet{linConstraintsGalacticOutflows2023}, shown as orange hatch
region in the top panel of Fig.~\ref{fig:ratio_estimate}. The cold gas mass,
which composes of neutral and molecular hydrogen, shown in the hatched region
in the middle panel comes from
\citet{guoNeutralUniverseMachineEmpiricalModel2023}. Taken together, we can
estimate the ratio of $T/\tau_{\rm eq}$ and the result is presented in the
bottom panel. This result is generally consistent from that obtained in \lgal:
$T/\tau \lesssim 20$ below $M_{\rm star}\approx 10^{10.5}\rm M_\odot$, and the
ratio increases for more massive galaxies. From these results, we conclude
that, even through galaxies within $M_{\rm star}\approx
10^{9.5-10.5}\rm M_\odot$ has entered the equilibrium state ($T\gtrsim
\tau_{\rm eq}$), stellar metallicity still has the memory from the past
history when the galaxy evolution is closer to the inflow-driven regime.
Therefore, variations in SFE, which is caused by different galaxy sizes
\citep{youngEfficiencyStarFormation1999, leroyStarFormationEfficiency2008,
shiExtendedSchmidtLaw2011, shiRevisitingExtendedSchmidt2018}, can still
imprint on the stellar metallicity, resulting in the anti-correlation between
galaxy size and stellar metallicity.

\section{Discussion}
\label{sec:Discussion}

\subsection{The role of $V_{\rm max}$}
\label{sub:the_role_of_v_rm_max_} % (fold)

\begin{figure}
  \begin{center}
    \includegraphics[width=0.95\linewidth]{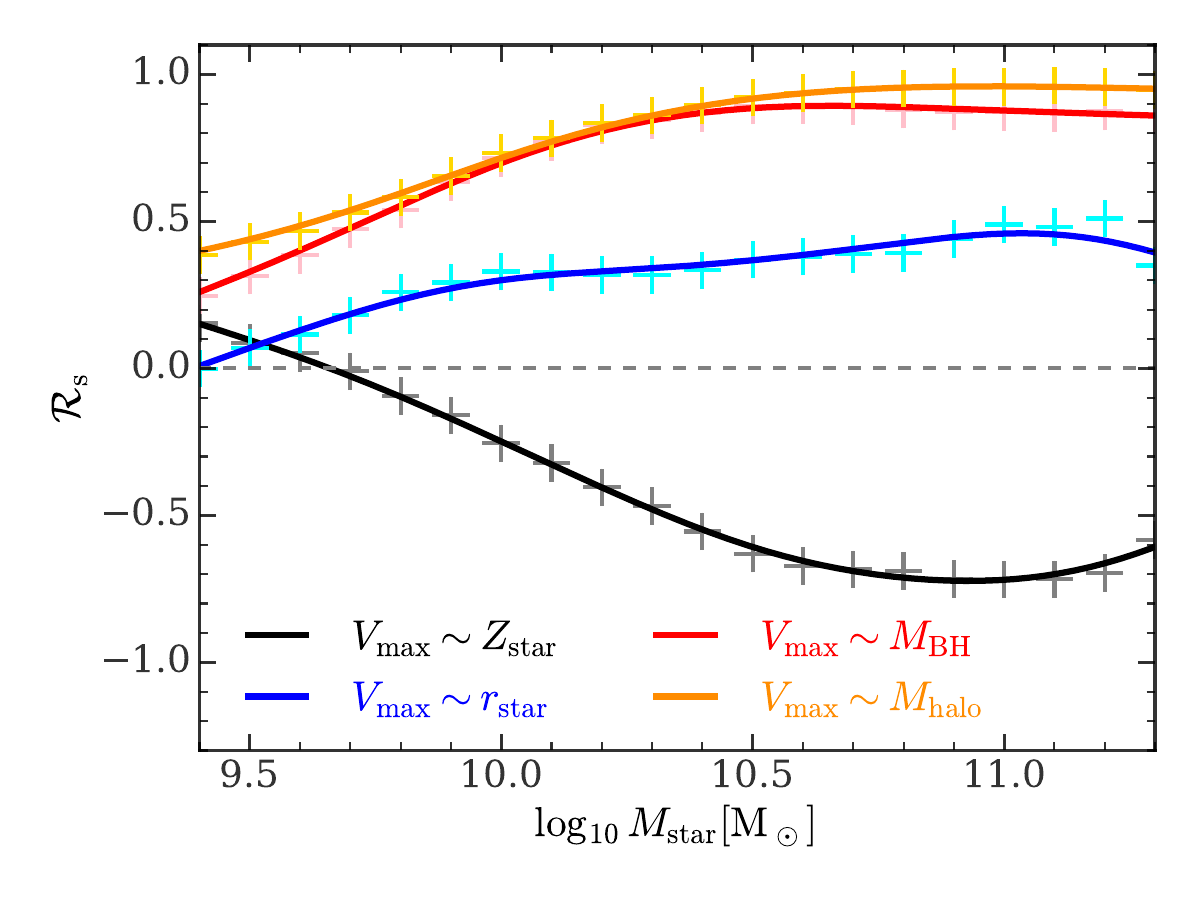}
  \end{center}
  \caption{
    Spearman's correlation coefficient between the maximum circular velocity,
    $V_{\rm max}$, and three galaxy properties: stellar metallicity ($Z_{\rm
    star}$), galaxy size ($r_{\rm star}$), and black hole mass ($M_{\rm BH}$).
  }
  \label{fig:vmax_corr}
\end{figure}

If SFE drives the variation in stellar metallicity, one might expect a strong
correlation between $V_{\rm max}$ and stellar metallicity in \lgal, as the
implemented SFE profoundly depends on $V_{\rm max}$ (see
Fig.~\ref{fig:lgal_sfe}). However, as shown in Fig~\ref{fig:vmax_corr}, the
correlation with $V_{\rm max}$ is weak at $M_{\rm star}\sim 10^{9.5}\rm
M_\odot$ and becomes negative at higher masses. Several effects contribute to
this result. First, $V_{\rm max}$ correlates with size, partially cancelling
its direct link to SFE. Second, $V_{\rm max}$ is tightly correlated with black
hole mass, so that galaxies with higher $V_{\rm max}$ are more likely to be
quenched and subsequent growth through mergers with lower-metallicity
satellites dilutes their metallicity. Third, higher $V_{\rm max}$ can enhance
SFE and potentially increase stellar metallicity, but it also corresponds to
higher halo mass and the additional accreted pristine gas can effectively
dilute the metal content and lower the gas-phase and stellar metallicity.
Together, these effects suppress the expected $V_{\rm max}$-stellar metallicity
relation and even invert it at the high-mass end. A full analysis of these
competing processes is beyond the scope of this work.

\subsection{Galaxy size-stellar metallicity correlation to high $z$
in L-GALAXIES}
\label{sub:galaxy_size_stellar_metallicity_correlation_to_high_z_in_l_galaxies}
% (fold)

\begin{figure}
  \begin{center}
    \includegraphics[width=0.95\linewidth]{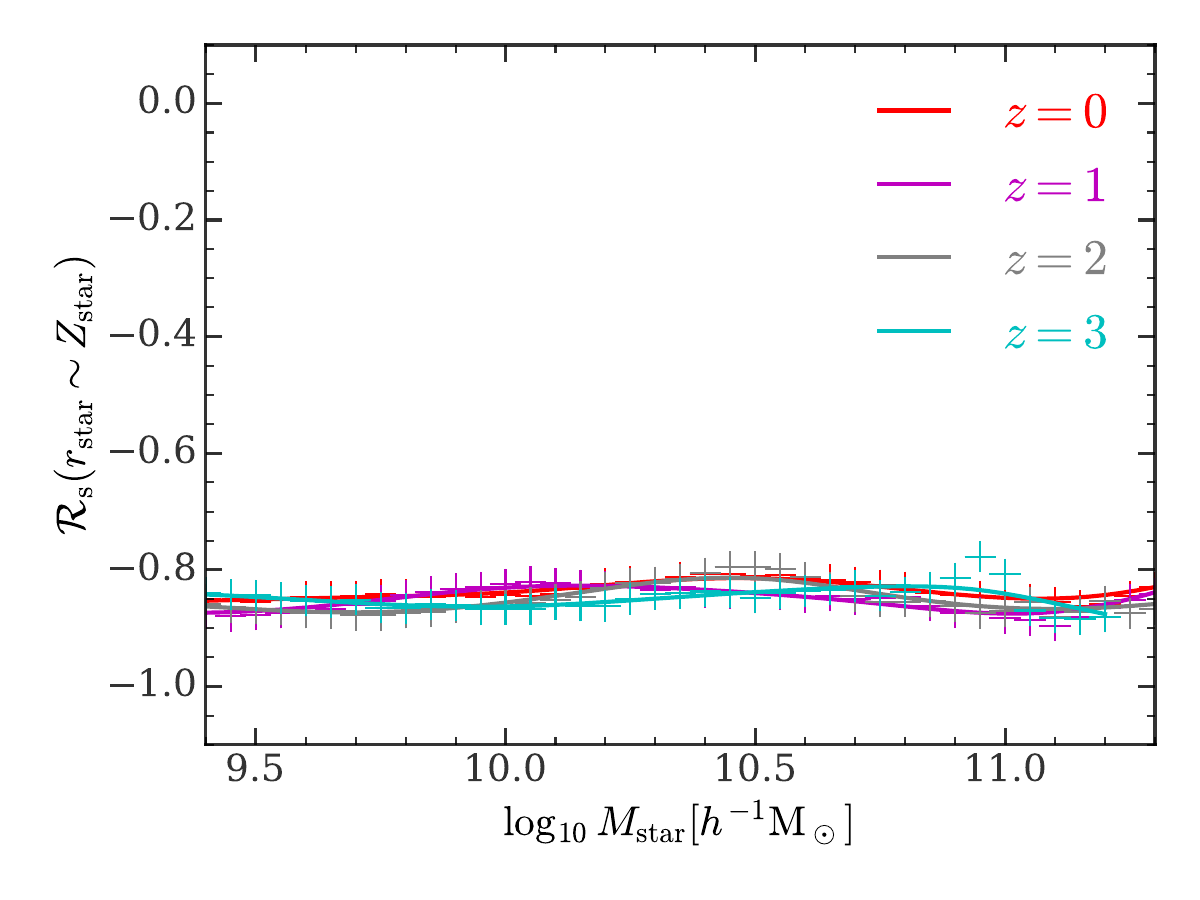}
  \end{center}
  \caption{
    Spearman's rank correlation coefficient between galaxy size and stellar
    metallicity as a function of stellar mass from $z=0$ to $z=3$ for central
    galaxies in \lgal. These two properties are manifesting a strong and
    statistically significant correlation ($\mathcal R_{\rm s}\approx
    -0.85$) for the whole stellar mass range we probed ($10^{9.5}-10^{11.5}\rm
    M_\odot$) across cosmic time.
  }
  \label{fig:correlation_redshift}
\end{figure}

To assess whether the size–stellar metallicity relation persists beyond the
local Universe, we extend our analysis to $z\sim 3$ in the \lgal model.
Fig.~\ref{fig:correlation_redshift} shows the Spearman correlation coefficients
between $r_{\rm star}$ and stellar metallicity for central galaxies from $z=0$
to $z=3$. Across the entire stellar mass interval
$10^{9.5}$--$10^{11.5}\,\mathrm{M_\odot}$, the anti-correlation remains strong,
with $\mathcal{R}_{\rm s}\simeq -0.8$ to $-0.85$ at all epochs.

The near constancy of the correlation amplitude suggests that the physical
origin of the size–metallicity relation is not a late-time phenomenon but is
instead imprinted throughout most of cosmic history in the model. In \lgal,
compact galaxies at fixed stellar mass form their stars more efficiently and
retain a larger fraction of their metals; this behaviour appears to hold at
both early and late times. The persistence of the relation therefore indicates
that star-formation efficiency and metal retention remain tightly linked to the
structural evolution of galaxies across cosmic time.

Although direct observational constraints on stellar metallicity and size at
$z>1$ remain limited, forthcoming high-redshift spectroscopic surveys such as
Subaru/PFS \citep{greenePrimeFocusSpectrograph2022} and VLT/MOONS
\citep{maiolinoMOONRISEMainMOONS2020} will be able to test these predictions
and provide an important benchmark for galaxy formation models.

\subsection{Size-metallicity relation for massive galaxies}
\label{sub:massive} % (fold)

A natural question is whether the size--metallicity anti-correlation persists
in the most massive galaxies, whose stellar growth is dominated by
\textit{ex-situ} mergers with lower-mass galaxies rather than \textit{in-situ}
star formation  \citep[e.g.][]{rodriguez-gomezStellarMassAssembly2016}. Both
the \manga data and the \lgal model indicate that the anti-correlation remains
strong at $M_{\rm star}\gtrsim 10^{11}\rm M_\odot$ (see
Figs.~\ref{fig:manga_correlation} and \ref{fig:lgal_correlation}).

The underlying mechanism differs from that at lower masses. For stellar
metallicity, dry mergers simply mix the pre-existing stellar populations: the
metallicity of the remnant is close to the mass-weighted average of the
progenitors. By contrast, the structural evolution is more dramatic. Analytic
arguments based on energy conservation predict that the size of a merger
remnant grows more rapidly than its stellar mass
\citep{coleHierarchicalGalaxyFormation2000}. Specifically,
\begin{equation}
  \frac{(M_1 + M_2)^2}{r_{\rm remnant}}\sim \frac{M_1^2}{r_1} +
  \frac{M_2^2}{r_2}
  + f_{\rm orbit}\frac{M_1M_2}{r_1 + r_2}
\end{equation}
where $M_1$, $M_2$, $r_1$, $r_2$ are the mass and size of progenitor galaxies,
respectively, $f_{\rm orbit}$ accounts for the orbital energy, and $r_{\rm
remnant}$ is the effective radius of the merger remnant. This estimate implies
a strong correlation between the size of the remnant galaxy and that of its
progenitor galaxies.

Taken together, these results imply that the anti-correlation between size and
stellar metallicity established at lower masses is inherited by massive
galaxies. Metallicities combine nearly linearly, while sizes grow
super-linearly in mergers. The outcome is that the negative size--metallicity
relation survives, and may even be amplified, in the most massive galaxies
where \textit{ex-situ} growth dominates.

\subsection{Implications to galaxy-halo connection}
\label{sub:Implications to galaxy-halo connection} % (fold)

\begin{figure}
  \begin{center}
    \includegraphics[width=0.95\linewidth]{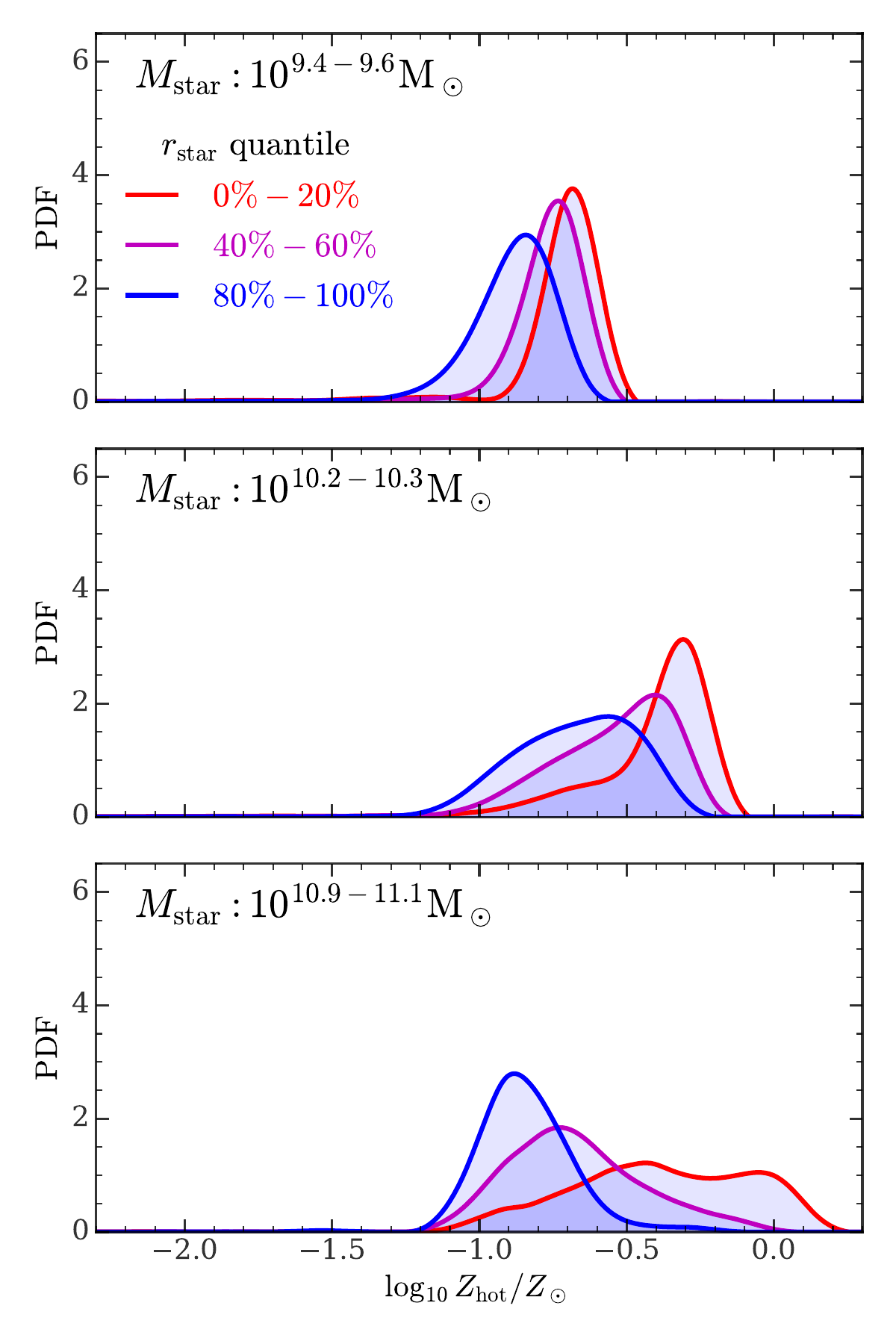}
  \end{center}
  \caption{
    The normalized distribution of hot gas stellar metallicity for central
    galaxies with different stellar mass (different panels) and different
    galaxy size ranks (different colors) in \lgal. At fixed stellar mass,
    central galaxies with smaller sizes can more efficiently pollute their hot
    gas atmosphere, thus their host haloes have more higher-metallicity hot gas.
  }
  \label{fig:mixing}
\end{figure}

\begin{figure}
  \begin{center}
    \includegraphics[width=0.95\linewidth]{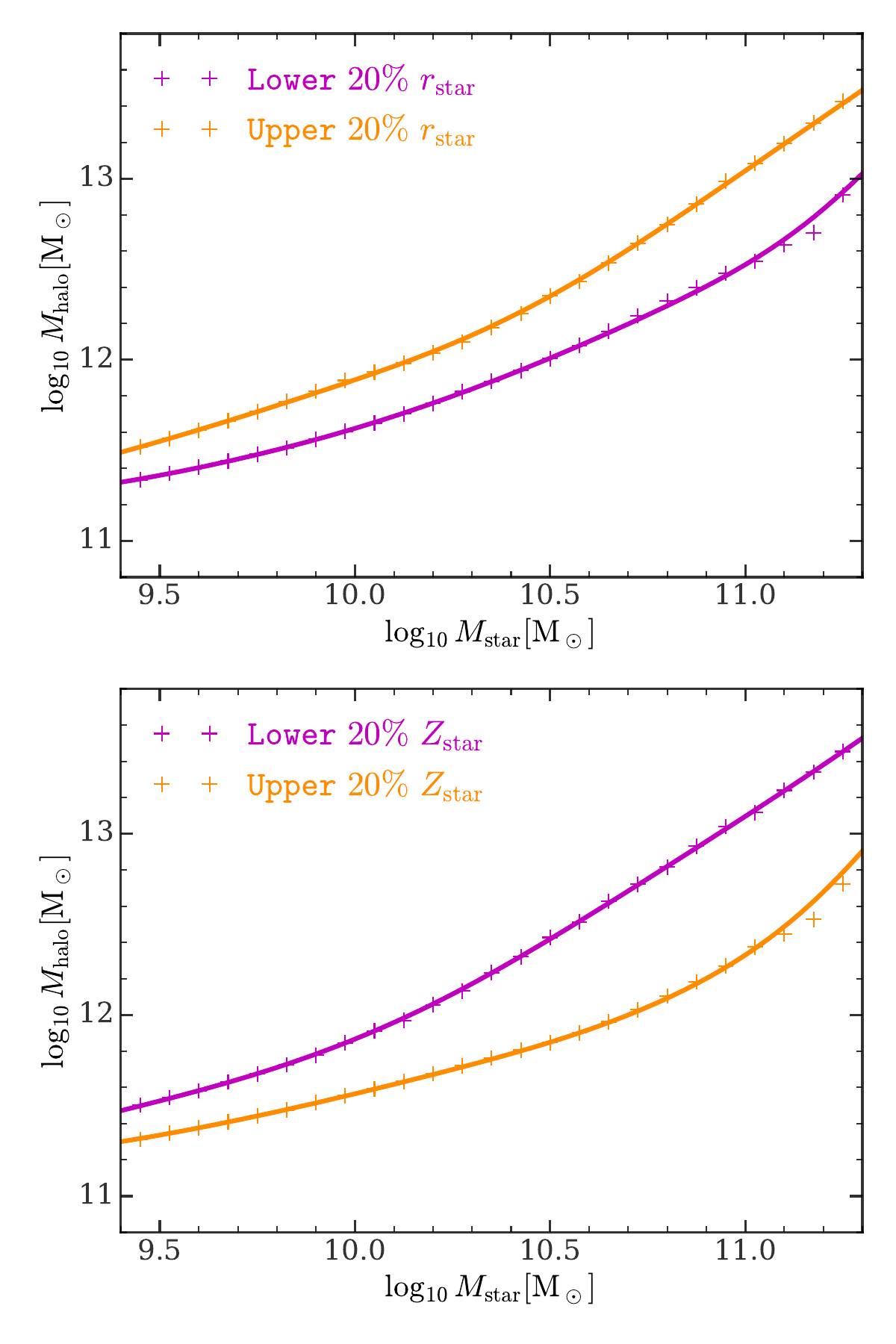}
  \end{center}
  \caption{
    The median halo mass as a function of stellar mass for central galaxies
    with different galaxy sizes (top panel) and stellar metallicity (bottom
    panel). At fixed stellar mass, central galaxies with larger sizes and lower
    stellar metallicity prefer to reside in more massive haloes than their
    star-forming counterparts.
  }
  \label{fig:shmr}
\end{figure}

We have seen that SFE regulates stellar metallicity primarily in the
inflow-driven regime ($T\ll \tau_{\rm eq}$), where SFE is low. This behaviour
is reflected in \lgal, where the size--metallicity relation is relatively
shallow (slope $\approx -0.3$) for compact galaxies ($r_{\rm star}\lesssim 2\rm
kpc$), but steepens (slope $\approx -0.6$) for larger systems, as shown in
Fig.~\ref{fig:lgal_correlation}. These slopes are steeper than predicted by the
gas regulator model under the assumption $\epsilon \propto r_{\rm star}^{-1}$,
which would yield $\approx -0.5$ at low SFE and flatten toward zero at high
SFE. The discrepancy indicates that processes beyond SFE contribute to the
galaxy size--stellar metallicity anti-correlation.

One contributing factor appears to be the metallicity of inflowing gas. At
fixed stellar mass, compact galaxies convert gas into stars more efficiently,
and thus typically reside in lower-mass haloes. For such systems the baryonic
reservoir is smaller, so the fixed stellar metal budget ($yM_{\rm star}/(1 -
R)$) enriches the recycled gas more effectively. This leads to systematically
more metal-rich inflows for compact galaxies. Fig.~\ref{fig:mixing} illustrates
this effect: the hot halo gas around compact galaxies is more metal-rich than
around extended galaxies of the same stellar mass. Recycling of enriched gas
likely enhance the galaxy size--stellar metallicity relation.

The link between size, SFE, and halo mass has further implications for the
stellar mass--halo mass (SMHM) relation. If smaller galaxies at fixed stellar
mass have higher SFE, they must inhabit lower-mass haloes than their extended
counterparts. Fig.~\ref{fig:shmr} confirms this expectation: at fixed stellar
mass, the smallest 20\% galaxies reside in haloes $\sim 0.2$ dex less massive
than the largest 20\% galaxies at $M_{\rm star}\sim 10^{9.5}\rm
M_\odot$, with the offset growing to $\sim 0.6$ dex by $M_{\rm star}\sim
10^{11}\rm M_\odot$. A similar trend appears when selecting galaxies by stellar
metallicity, consistent with metallicity tracing SFE. These offsets represent a
direct, testable prediction: central galaxies of the same stellar mass but
different size or metallicity should occupy haloes of systematically different
mass.

Observational evidence already points in this direction:
\citet{charltonDependenceHaloMass2017}, using weak gravitational lensing, found
that at fixed stellar mass, larger galaxies tend to reside in more massive
haloes, which is consistent with \lgal results. Further confirmation with
larger lensing samples or kinematic halo mass estimates would provide a
stringent test of this prediction and help disentangle the roles of SFE and
recycling. More generally, the connection between size, stellar metallicity,
and halo mass highlights how galaxy structure encodes information about the
baryon cycle and its coupling to dark matter haloes.

\subsection{Relation to central galaxy quenching}
\label{sub:relation_to_galaxy_quenching} % (fold)

An important question is how the galaxy size--stellar metallicity relation
relates to the quenching of central galaxies
\citep{pengMassEnvironmentDrivers2010}. While the physical drivers of central
galaxy quenching remain debated, a number of empirical trends are
well-established: at fixed stellar mass, quiescent central galaxies tend to be
more compact \citep[e.g.][]{kauffmannStellarMassesStar2003,
  brinchmannPhysicalPropertiesStarforming2004, wooTwoConditionsGalaxy2015,
lillySURFACEDENSITYEFFECTS2016, genelSizeEvolutionStarforming2018}, more
metal-rich \citep[e.g.][]{pengStrangulationPrimaryMechanism2015,
trusslerBothStarvationOutflows2020}, and reside in more massive haloes
\citep[e.g.][]{mandelbaumStrongBimodalityHost2016, zuMappingStellarContent2016,
zhangMassiveStarformingGalaxies2022, wangTestingGalaxyFormation2025}. Within
our framework, the first two trends arise naturally: quiescent central galaxies
must have formed their stars more efficiently in the past, requiring higher gas
surface densities and thus smaller sizes, and it leads to rapid enrichment and
higher stellar metallicities. However, the trend in halo mass appears more
puzzling. Because galaxies with higher SFE tend to reside in lower-mass haloes
at fixed stellar mass, one would expect quenched galaxies, which are compact
and metal-rich, to occupy less massive haloes. A natural resolution is that
halo growth continues unabated after quenching, and eventually these quiescent
galaxies tend to reside in more massive haloes
\citep{pengMASSENVIRONMENTDRIVERS2012, wangTestingGalaxyFormation2025}. In this
scenario, quiescent central galaxies retain the compact sizes and high
metallicities imprinted by their efficient early growth, while subsequent halo
accretion allows them to surpass the halo masses of their star-forming
counterparts.

This scenario also suggests a close link between central galaxy quenching and
the assembly history of their host haloes. Quiescent central galaxies, having
formed stars more efficiently early on, are expected to reside  in relatively
low-mass haloes at that time. Then subsequent halo growth builds up more
massive haloes around them compared to their star-forming counterparts. This
picture implies that quiescent central galaxies preferentially occupy
late-forming haloes, while star-forming central galaxies reside in haloes that
assembled earlier. The link between central galaxy quenching and halo assembly
is consistent with observations by \citet{wangLateformedHaloesPrefer2023}, who
used the stellar mass-to-halo mass ratio as a proxy for halo formation time and
found that quiescent central galaxies tend to inhabit late-formed haloes
\citep[see also][]{cuiOriginGalaxyColour2021, moTwophaseModelGalaxy2024,
wangTestingGalaxyFormation2025}.

\subsection{Caveats}
\label{sub:caveats} % (fold)

Although semi-analytical models allow us to isolate the impact of individual
physical processes due to their explicit prescription. This also limits their
capability to model galaxy evolution realistically. Firstly, \lgal does not
include baryon's self-gravity and the corresponding halo response into the
gravitational potential calculation
\citep[e.g.][]{blumenthalContractionDarkMatter1986,
gnedinResponseDarkMatter2004, bensonGalaxyFormationSpanning2010}, and these
effects can deepen the potential and reduce galaxy sizes, thus strengthening
the correlation between galaxy size and gravitational potential. Secondly,
effective feedback not only needs to escape the gravitational potential, but
also needs to survive the hydrodynamical interaction of surrounding
interstellar medium, during which process the energy can be transferred to
surrounding interstellar medium and radiated away. Meanwhile, galaxy size
traces the density of interstellar medium, which indicates the strength of this
effect. Therefore, the correlation between galaxy size and stellar metallicity
may still be contributed by the difference in gravitational and
non-gravitational potential that affects the effectiveness of stellar feedback.
Nonetheless, our work here clearly shows that SFE plays a non-negligible role
in both semi-analytical models and our real Universe.

\section{Summary}
\label{sec:Summary}

The metallicity of galaxies carries crucial information on the interplay of
physical processes during galaxy formation and evolution. While the gas-phase
metallicity probes processes on shorter timescales, stellar metallicity is like
the fossil record of the whole galaxy evolution history. For gas-phase
metallicity, smaller galaxies are known to have higher metallicities than their
more extended counterparts at fixed stellar mass, and it has been attributed to
deeper gravitational potential and temporally correlated star formation
histories. Extending this line of inquiry to stellar metallicity, we find
similar anti-correlation for central galaxies in our local Universe using
\manga dataset. By investigating the analogous relation in \lgal, we rule out
gravitational potential and star formation history as the primary drivers, and
turn instead to SFE. We demonstrate the connection between SFE and stellar
metallicity using a simple gas-regulator framework, which shows that SFE
affects stellar metallicity only when the effective galaxy evolution time $T$
is not much longer than the equilibrium timescale $\tau_{\rm eq}$, or
$T/\tau_{\rm eq}\ll 20$. Further we show that, both in observation and \lgal,
galaxies below $10^{10.5}\rm M_\odot$ are within this regime. Finally, we
speculate that massive galaxies can still maintain this anti-correlation
because galaxy merging process preserve the relative rank of galaxy size and
stellar metallicity. Our main results can be summarized as follows:
\begin{itemize}

  \item Using the \manga dataset, we find a clear anti-correlation between
    stellar metallicity and stellar half-mass radius at fixed stellar mass for
    central galaxies (see Figs.~\ref{fig:manga_distribution} and
    \ref{fig:manga_correlation}). The correlation becomes stronger with stellar
    mass, from $\mathcal R_{\rm s} \approx -0.2$ at $M_{\rm star}\sim
    10^{9.5}\rm M_\odot$ to $\mathcal R_{\rm s} \approx -0.4$ at $M_{\rm
    star}\sim 10^{10}\rm M_\odot$ and above.

  \item \lgal reproduces this trend, with a stronger correlation ($\mathcal
    R_{\rm s}\approx -0.8$; Fig.~\ref{fig:lgal_correlation}). The slope of the
    galaxy size-metallicity relation is $\approx -0.3$ for compact galaxies
    ($r_{\rm star}\lesssim 2,\rm kpc$) and steepens to $\approx -0.6$ at larger
    radii.

  \item Controlling for gravitational potential depth ($V_{\rm max}$) and
    stellar age does not eliminate the correlation (see
    Fig.~\ref{fig:lgal_vmax_age}), indicating that it is not driven by feedback
    retention or by recent inflows. Instead, the correlation arises from
    variations in star formation efficiency, which systematically decreases
    with galaxy size (Fig.~\ref{fig:lgal_sfe}).

  \item The analytical gas-regulator model shows that in the inflow-driven
    regime ($T\ll \tau_{\rm eq}$), stellar metallicity scales with the square
    root of SFE, i.e. $\langle Z_{\rm star}\rangle \propto \sqrt{\epsilon}$,
    while in equilibrium it saturates at a level determined by the metallicity
    of the inflowing gas (see Fig.~\ref{fig:model}).

  \item We estimate the ratio between galaxy evolution time and the equilibrium
    timescale, $T/\tau_{\rm eq}$, and find that galaxies with $M_{\rm
    star}\lesssim 10^{10.5}$ remain the regime where SFE influences stellar
    metallicity (see Fig.~\ref{fig:ratio_estimate}).

  \item In \lgal, the relation between stellar metallicity and $V_{\rm max}$ is
    more complex. The correlation is weak at $M_{\rm star}\sim 10^{9.5}\rm
    M_\odot$ and turns negative at higher masses (see
    Fig.~\ref{fig:vmax_corr}). This likely reflects the combined influence of
    correlations with galaxy size, black hole mass, and halo mass, which
    together weaken---and at higher masses even invert---the expected trend.

  \item The anti-correlation between galaxy size and stellar metallicity
    persists at $M_{\rm star}\gtrsim 10^{11},\rm M_\odot$, where the growth of
    galaxies is dominated by mergers, as the size and stellar metallicity of
    the remnant galaxy depend strongly on that of their progenitor galaxies
    (see \S\,\ref{sub:massive}).

  \item The enrichment of hot halo gas depends on galaxy size, where smaller
    galaxies have more enriched hot gas than their more extended counterparts
    (see Fig.~\ref{fig:mixing}). The stellar mass-halo mass relation is
    dependent on galaxy size and stellar metallicity: at fixed stellar mass,
    galaxies with the lowest 20\% sizes reside in haloes $\approx 0.2$ dex at
    $M_{\rm star} \sim 10^{9.5}\rm M_\odot$, and $\approx 0.6$ dex at $M_{\rm
    star}\approx 10^{11}\rm M_\odot$, less massive than galaxies with the
    highest 20\% sizes. A similar offset is found for stellar metallicity,
    where central galaxies with lower stellar metallicities live in more
    massive haloes (see Fig.~\ref{fig:shmr}).

\end{itemize}

The galaxy size--stellar metallicity relation thus provides a new observational
window into the baryon cycle. By linking stellar metallicity to both the
structural growth and gas processing, it provides a simple but powerful
diagnostic of how galaxies evolve. Our results establish the relation for
central galaxies in our local Universe and identify SFE and metal recycling as
key drivers. Future weak gravitational lensing measurements and hydrodynamical
simulations will help to confirm this picture and clarify the extent to which
it holds across different environments and cosmic times.

\section*{DATA AVAILABILITY}

The data supporting the plots within this article are available on reasonable
request to the corresponding author.

\section*{Acknowledgements}

The author thanks the anonymous referee for their helpful comments that
improved the quality of the manuscript. KW thanks Yingjie Peng, Enci Wang, N.
F. Boardman, Peder Norberg, Sownak Bose, and Yangyao Chen for helpful
discussions. KW acknowledges support from the Science and Technologies
Facilities Council (STFC) through grant ST/X001075/1.

This work used the DiRAC@Durham facility managed by the Institute for
Computational Cosmology on behalf of the STFC DiRAC HPC Facility
(www.dirac.ac.uk). The equipment was funded by BEIS capital funding
via STFC capital grants ST/K00042X/1, ST/P002293/1, ST/R002371/1 and
ST/S002502/1, Durham University and STFC operations grant
ST/R000832/1. DiRAC is part of the National e-Infrastructure.

The computation in this work is supported by the HPC toolkit \specialname[HIPP]
\citep{hipp}, IPYTHON \citep{perezIPythonSystemInteractive2007}, MATPLOTLIB
\citep{hunterMatplotlib2DGraphics2007}, NUMPY
\citep{vanderwaltNumPyArrayStructure2011}, SCIPY
\citep{virtanenSciPyFundamentalAlgorithms2020}, ASTROPY
\citep{robitailleAstropyCommunityPython2013}.
This research made use of NASA’s Astrophysics Data System for bibliographic
information.

\bibliographystyle{mnras}
\bibliography{bibtex}

\appendix

\section{Estimating the effective loading factor}
\label{sec:estimating_the_effective_loading_factor}

\citet{luAnalyticalModelGalaxy2015} proposed an analytical method to estimate
the mass loading factor based on book-keeping the metal flow. It is based on
two differential equations that governs the metal mass evolution in the
interstellar medium
\begin{equation}
  \frac{\dd (Z_{\rm gas}M_{\rm gas})}{\dd t} = - (1 - R  + \eta_{\rm
  eff})Z_{\rm gas}\Psi + y\Psi
\end{equation}
and in long-lived stars
\begin{equation}
  \frac{\dd (Z_{\rm star}M_{\rm star})}{\dd t} = Z_{\rm g}\frac{\dd
  M_{\rm star}}{\dd t}
\end{equation}
where $\Psi = (\dd M_{\rm star} /\dd t)/(1 - R)$ is the star formation rate.
Here $\eta_{\rm eff}$ is the loading factor, which is subject the depth of the
gravitational potential, so it should be closely related to stellar mass and
evolve as the galaxy grows. Here we regard it as an average over the whole
evolution history and treat it as a constant. Combined these two equations, we
have
\begin{equation}
  \frac{\dd (Z_{\rm gas}M_{\rm gas})}{\dd t} = -\frac{1 - R +
  \eta_{\rm eff}}{1 -
  R}\frac{\dd (Z_{\rm star}M_{\rm star})}{\dd t} + \frac{y}{1- R}\frac{\dd
  M_{\rm star}}{\dd t}
\end{equation}
and it integrates into
\begin{equation}
  Z_{\rm gas}M_{\rm gas} = -\frac{1 - R + \eta_{\rm eff}}{1 -
  R}Z_{\rm star}M_{\rm star} + \frac{y}{1- R}M_{\rm star}
\end{equation}
which can be rearranged into
\begin{equation}
  \eta_{\rm eff} = (1 - R) \left[\frac{y}{(1 - R)Z_{\rm star}} - \frac{Z_{\rm
  gas}M_{\rm gas}}{Z_{\rm star}M_{\rm star}} - 1\right]
\end{equation}

\section{Galaxy size-stellar metallicity relation in \lgal}
\label{sec:galaxy_size_stellar_metallicity_relation_in_lgal}

\begin{figure*}
  \begin{center}
    \includegraphics[width=0.95\linewidth]{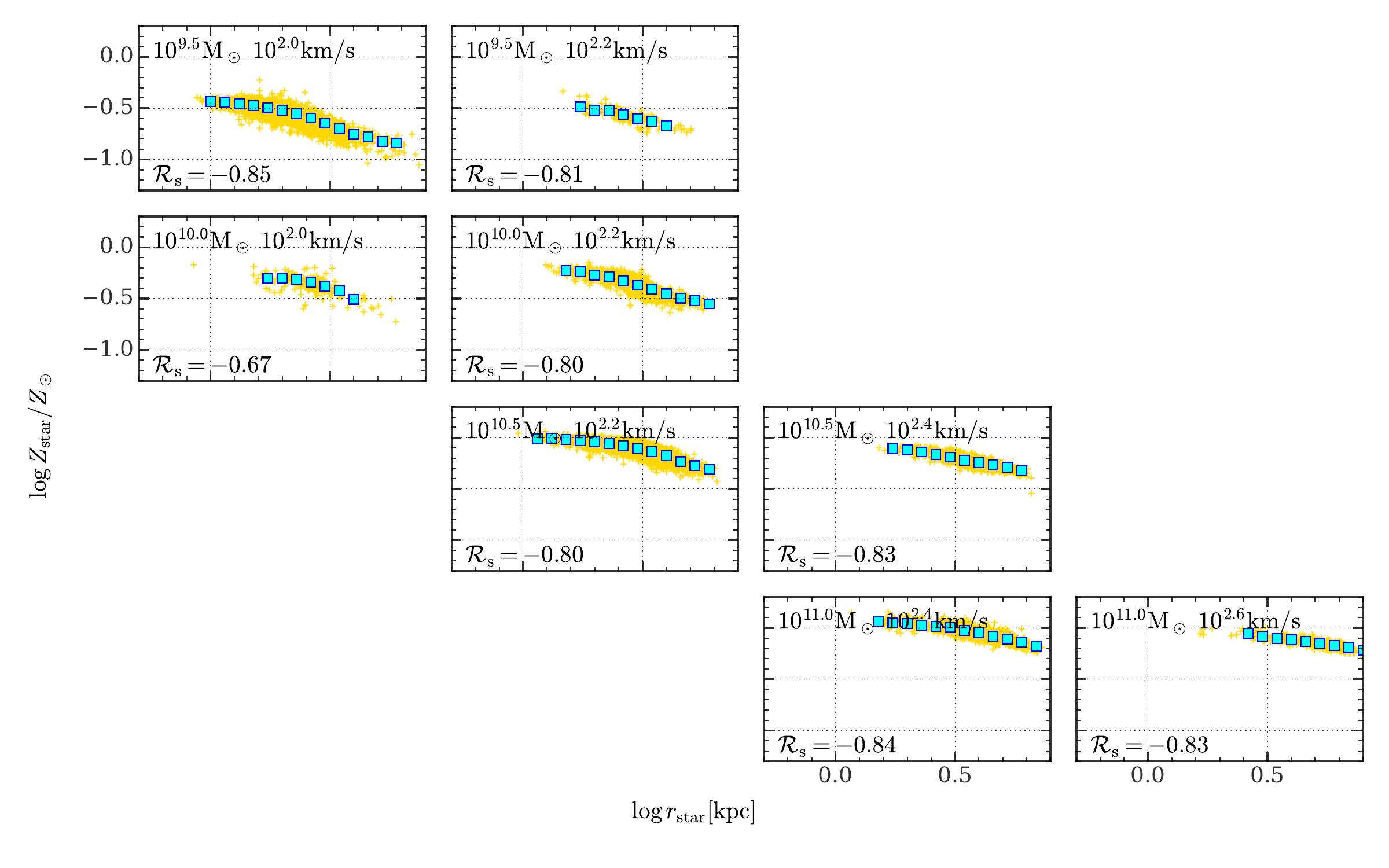}
  \end{center}
  \caption{
    The joint distribution of galaxy size and stellar metallicity in fine bins
    of stellar mass and $V_{\rm max}$ (0.1 dex). The cyan boxes show the median
    stellar metallicity in bins of galaxy size. Here only bins with more than
    30 galaxies are shown.
  }
  \label{fig:joint_control_vmax}
\end{figure*}

\begin{figure*}
  \begin{center}
    \includegraphics[width=0.95\linewidth]{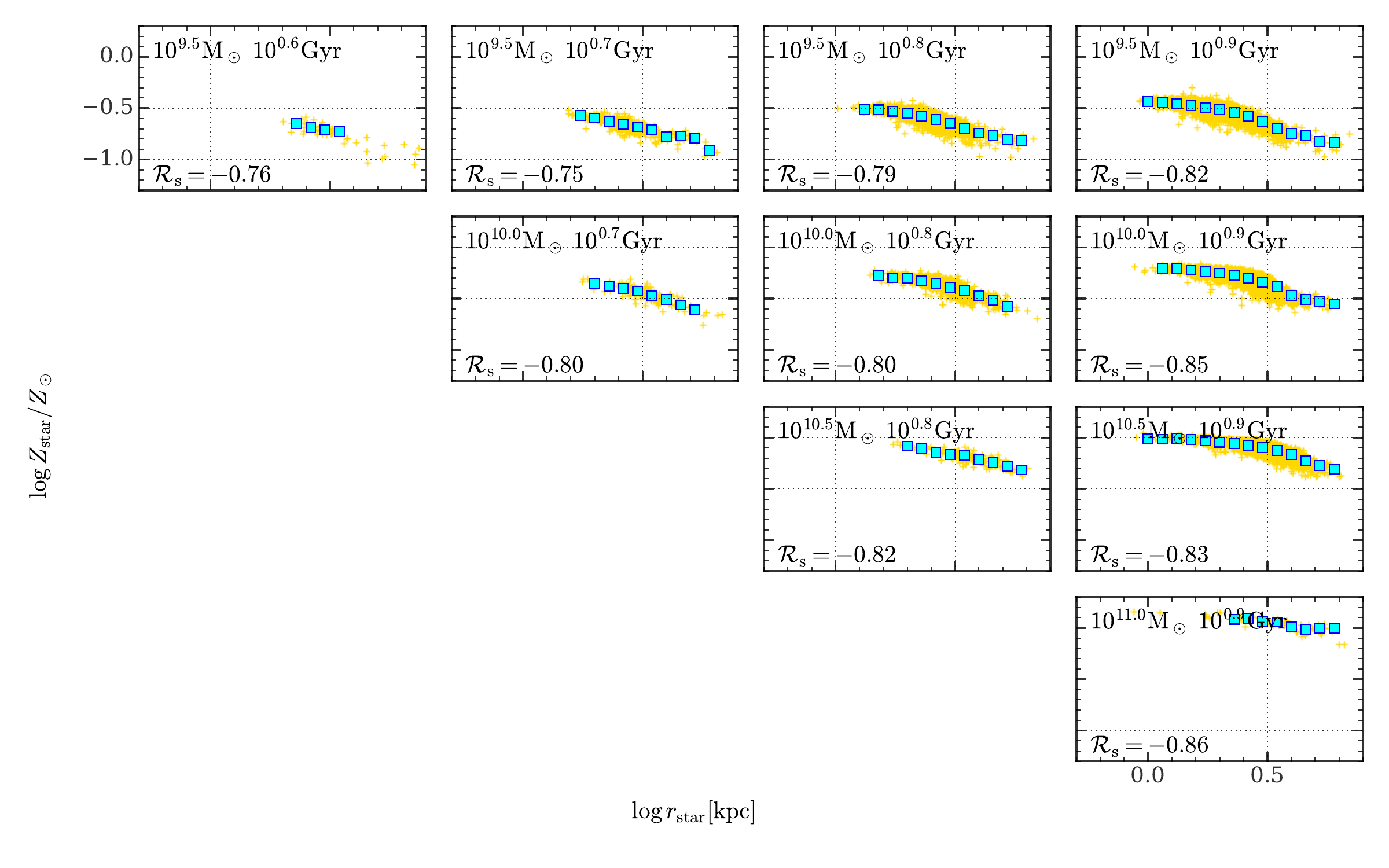}
  \end{center}
  \caption{
    The joint distribution of galaxy size and stellar metallicity in fine bins
    of stellar mass and stellar age(0.1 dex). The cyan boxes show the median
    stellar metallicity in bins of galaxy size. Here only bins with more than
    30 galaxies are shown.
  }
  \label{fig:joint_control_age}
\end{figure*}

Figs.~\ref{fig:joint_control_vmax} and \ref{fig:joint_control_age} show the
joint distribution of galaxy size and stellar metallicity in fine bins of
$V_{\rm max}$ and stellar age, both with stellar mass fixed. The correlation
strength between galaxy size and stellar metallicity is not affected by $V_{\rm
max}$, nor stellar age. Moreover, the median relation between stellar
metallicity and galaxy size are similar to each other in different bins of
$V_{\rm max}$ and stellar age, with stellar mass fixed.

% Don't change these lines
\bsp  % typesetting comment
\label{lastpage}
\end{document}